\documentclass[aps,preprint,nofootinbib]{revtex4-1}

\usepackage[utf8]{inputenc}
\usepackage{amsfonts}
\usepackage{color} 

\usepackage{graphicx}
\usepackage{subfig}
\usepackage{showlabels}
\usepackage{amssymb}
\usepackage{amsmath}
\usepackage{stmaryrd}
\usepackage{latexsym}
\usepackage{multirow}

\usepackage{caption}
\usepackage{xcolor}
\definecolor{myurlcolor}{HTML}{08457E}
\definecolor{mylinkcolor}{HTML}{2A52BE}
\definecolor{mycitecolor}{HTML}{E30022}

\usepackage{float}
\usepackage[colorlinks, linkcolor=mylinkcolor, citecolor=mycitecolor, urlcolor=myurlcolor, linktocpage=true]{hyperref}

\def\equationautorefname~#1\null{(#1)\null}
\def\tableautorefname~#1\null{(#1)\null}
\def\figureautorefname~#1\null{(#1)\null}
\def\sectionautorefname~#1\null{(#1)\null}
\def\subsectionautorefname~#1\null{(#1)\null}

\let\origref\autoref
\def\autoref#1{\textbf{\origref{#1}}}

\let\origcite\cite
\def\cite#1{\textbf{\origcite{#1}}}

\usepackage{titlesec}
\titleformat*{\section}{\centering\small\bfseries\scshape}
\titleformat*{\subsection}{\small\bfseries\scshape}
\titleformat*{\subsubsection}{\small\bfseries\scshape}

\newcommand{\be}{\begin{equation}}
\newcommand{\ee}{\end{equation}}
\newcommand{\bea}{\begin{eqnarray}}
\newcommand{\eea}{\end{eqnarray}}
\newcommand{\benn}{\begin{eqnarray*}}
\newcommand{\eenn}{\end{eqnarray*}}

\def\bse{\begin{subequations}}%
\def\ese{\end{subequations}}%

\def\R{{\cal R}}     

\newcommand{\sfrac}[2]{\dfrac{\,#1\,}{\,#2\,}}

\renewcommand{\d}{\mathrm{d}}
\newcommand{\der}[2]{\sfrac{\d #1}{\d #2}}

\newcommand{\tl}[1]{\tilde{#1}}
\newcommand{\m}{m_{\mathrm{Pl}}}

\newcommand{\SReps}{\text{\large{$\epsilon$}}}
\newcommand{\SReta}{\text{\large{$\eta$}}}

\let\oldsqrt\sqrt
\def\sqrt{\mathpalette\DHLhksqrt}
\def\DHLhksqrt#1#2{%
	\setbox0=\hbox{$#1\oldsqrt{#2\,}$}\dimen0=\ht0
	\advance\dimen0-0.4\ht0
	\setbox2=\hbox{\vrule height\ht0 depth -\dimen0}%
	{\box0\lower0.4pt\box2}}

\def\prd{Physical Review D}

\def\plb{Physics Letters B}
\def\pla{Physics Letters A}

\begin{document}

\title{Formalizing Slow-roll Inflation\\
in \\
Scalar-Tensor Theories of Gravitation \vspace{5mm}}

\author{Kemal Akın}
\email{akinkem@itu.edu.tr}

\author{A.\ Savaş Arapoğlu}
\email{arapoglu@itu.edu.tr}

\author{A.\ Emrah Yükselci}
\email{yukselcia@itu.edu.tr}

\affiliation{\vspace{5mm}Istanbul Technical University, Faculty of Science and Letters, Physics Engineering Department, 34469, Maslak, Istanbul, Turkey \vspace{2cm}}

\begin{abstract}
The viability of slow-roll approximation is examined by considering the structure of phase spaces in scalar-tensor theories of gravitation and the analysis is exemplified with a nonminimally coupled scalar field to the spacetime curvature. The slow-roll field equations are obtained in the Jordan frame in two ways: first using the direct generalization of the slow-roll conditions in the minimal coupling case to nonminimal one, and second, conformal transforming the slow-roll field equations in the Einstein frame to the Jordan frame and then applying the generalized slow-roll conditions. Two inflationary models governed by the potentials $V(\phi) \propto \phi^2$ and $V(\phi) \propto \phi^4$ are considered to compare the outcomes of two methods based on the analysis of $n_s$ and $r$ values in the light of recent observational data.

\end{abstract}

\maketitle
\raggedbottom

\section{INTRODUCTION\label{sec:intro}}

Inflation is the most plausible scenario providing not only a successful explanation of the horizon, flatness, and monopole problems of the standard big bang cosmology \cite{guth81,linde82,alb-stein82}, but also the primordial density fluctuations for the formation of the observed large-scale structure of the universe (Refs.\ \cite{RMP97, PR99, RMP06, baumann09} for reviews).

In most inflationary universe models, it is supposed that the nearly exponential expansion of the universe is driven by a scalar field (called inflaton) which is assumed to be minimally coupled to the gravity and \textit{slowly} evolves in a nearly flat potential $V(\phi)$. In the so-called ``slow-roll (SR) approximation'' \cite{sr94} the most slowly changing terms in the field equations are neglected which amounts to the approximation that the kinetic energy of the inflaton is considered to be much smaller than  the potential energy, that is, $\dot{\phi }^{2 }\ll V(\phi )$ and $\ddot{\phi }\ll H\dot{\phi }$.  The single-field inflationary models predict almost scale-invariant density perturbations consistent with the observations of anisotropies in Cosmic Microwave Background (CMB).  But the existence of inflationary attractors is necessary for the SR approximation to work \cite{faraoni-gsr}.

On the other hand, quantum field theory in curved spacetime necessitates a non-trivial coupling between the scalar field and the spacetime curvature even if they are absent in the classical theory. Actually there are many other indications that the inflaton couples to the curvature of spacetime $\R$ (summarized in a nice way in Ref.\ \cite{faraoni96}). Therefore, it is reasonable to consider how the dynamics of the inflaton changes because of this nonminimal coupling. In general, one expects that the coupling is of the form $\xi \phi^2 \R$ with a constant $\xi $, but the quantum corrections may change this situation and the behaviour of renormalization group effective coupling $\xi$ becomes $\phi$ dependent also. Recently in this direction the running of the non-minimal parameter $\xi$ is analyzed within the non-perturbative setting of the functional renormalization group \cite{shapiro15} and the inflationary parameters in the renormalization group improved $\phi^{4}$ theory at one-loop and two-loop levels are considered in Refs.\ \cite{odintsov14,odintsov15}.  To cover all these effective models, then, one can consider a nonminimally coupled inflaton field with a general coupling function of the form $F(\phi)$ \cite{myrzakul15, myrzakul15a, zubair17, carsten16, he19, granda1, granda2, granda3, taka2003}.    

We are now currently in an era stated commonly as the ``precision cosmology", implying that the observational data sharpens and this allows one to compare the models more precisely. Inflationary models are examined and compared by the 2018 release of the Planck CMB anisotropy measurements \cite{planck2018}, by checking the inflationary parameters such as the spectral index $n_s$, the tensor-to-scalar ratio $r$, and the analysis is performed by the help the tools developed in Ref.\ \cite{Lewis2019_getdist}. Indeed, discriminating the various inflationary models through the calculation of these parameters in both minimally and nonminimally coupled theories is an active research area. Therefore, it is beneficial to consider and to compare the calculation of these parameters in nonminimally coupled theories, and to check the significance of difference between minimal and nonminimal cases, considering the recent bunch of papers appearing in the literature about the subject. Thus, the aim is to consider the inflation in the Jordan Frame (JF), without mapping into the Einstein Frame (EF) via conformal transformations and without discussing the equivalence of two frames or which frame is physical \cite{chiba2003, chiba13, sasaki10, sasaki11, kamen15, ohta17, ruf18, kuusk16, kannike17, tamvakis17}.  

The inflationary parameters, $n_s$ and $r$, are obtained in the SR approximation either considered directly in the JF through the ``generalized slow-roll'' (GSR) approximation \cite{morris01, torres97}, or by performing a conformal transformation to the EF and using the usual definitions of SR parameters \cite{sr94} in this frame; mostly the latter is preferred because of simplicity. The existence of attractor behaviour in inflation with nonminimal coupling is also demonstrated in Ref.\ \cite{faraoni-gsr} which is necessary for SR approximation to work. 

In this paper, the SR field equations are obtained in the JF in two ways. In the first method, the so-called ``generalized'' SR conditions are used directly in the JF, \cite{torres97, morris01}, and the SR field equations are given without any reference to the EF. In the second method, the SR field equations are written in the EF, as they are originally suggested, and the corresponding ones in the JF are obtained via conformal transformations followed by the GSR approximation. The aim of the paper is not to compare the calculation of any inflationary parameter in the JF and the EF but to get the SR field equations in the JF in a systematic way. There is an interesting difference between the two methods. Although the SR Friedmann equations coincide, the scalar field equations do not match exactly which leads to a difference in the calculation of inflationary parameters, the spectral index $n_s$ and the tensor-to-scalar ratio $r$.

The plan of the paper is as follows: The main equations, which are used throughout this study, and the notation are set in Sec.\ \autoref{sec:setup}. In Sec.\ \autoref{sec:viability_SR} the viability of SR approach is investigated via the dynamical system analysis after a brief discussion on the observational predictions of minimally coupled scalar field model. In Sec.\ \autoref{sec:SR_Eqs} the SR approximated equations of motion in the JF are obtained with two different methods mentioned above and the results are compared by calculating the inflationary parameters, $n_s$ and $r$. Then, the nonminimally coupled scalar field model is considered in Sec.\ \autoref{sec:example} as an example of the formal examination. Finally, the concluding remarks are given in Sec.\ \autoref{sec:conclusion}.

\section{SET-UP AND NOTATION \label{sec:setup}}

The action for the nonminimally coupled scalar-tensor theories in the JF is
\begin{equation}
    S_{\rm JF} = \int \d^4 x \, \sqrt{|g|} \left[ F(\phi) \R  - \sfrac{1}{2} g^{\mu \nu} \, \nabla_{\!\mu}\,\phi \, \nabla_{\!\nu}\,\phi - V(\phi) \right]
\label{eq:JFaction}
\end{equation}
where $F(\phi)$ and $V(\phi)$ are the coupling function and the potential of the scalar field, respectively. Considering the flat Friedmann-Lemaître-Robertson-Walker (FLRW) metric as
\begin{equation}
    \d s^2 = -\d t^2 + a^2(t) \big[ \d r^2 + r^2 \big( \d \theta^2 + \sin^2\!\theta  \,\d \varphi^2 \big) \big]
\label{eq:metric}
\end{equation}
the equations of motion obtained from the above action yield
\begin{subequations}
\begin{align}
    &6F(\phi)H^2 = \sfrac{1}{2} \dot{\phi}^2 + V(\phi) - 6 H \dot{F}(\phi) \:, \label{eq:JF_Friedmann} \\[2mm]
    &4F(\phi)\dot{H} = -\dot{\phi}^2 - \ddot{F}(\phi) + 2H\dot{F}(\phi) \:, \label{eq:JF_Acceleration} \\[2mm]
    &\ddot{\phi} + 3H\dot{\phi} - 6 \big(2H^2 + \dot{H}\big) F'(\phi) + V'(\phi) = 0 \label{eq:JF_Scalar}
\end{align}
\label{eq:Friedmann_scalar_eqs}%
\end{subequations}
where an overdot and a prime represent derivatives with respect to time and the scalar field, respectively. Additionally, the equation of state parameter is
\begin{equation}
    \omega_\phi = \sfrac{p_\phi}{\rho_\phi} = -1 - \sfrac{2\dot{H}}{3H^2}
\end{equation}
with the definitions of the density $\rho_\phi=3\m^2H^2$ and the pressure $p_\phi=-\m^2(2\dot{H}+3H^2)$ for the scalar field.\footnote{Throughout this study we set $c=\hbar=1$ and, consequently, the reduced Planck mass becomes $\m^2 \equiv 1/(8\pi G)$.}

Nevertheless, in nonminimally coupled inflationary models the method followed mostly in the literature is to map the model in the JF via conformal transformations to a model in the EF, presumably due to the fact that the field equations and the procedure to follow are simpler in the EF in comparison with the JF. Therefore, performing a conformal transformation of the form
\begin{equation}
    \tl{g}_{\mu \nu} = \Omega^2(\phi)\,g_{\mu \nu} \:, \qquad \Omega^2(\phi) = \sfrac{2}{\m^2} F(\phi)
\label{eq:CT}
\end{equation}
the action in the JF given in Eq.\ \autoref{eq:JFaction} becomes in the EF
\begin{equation}
    S_{\rm EF} = \int \d^4 x \, \sqrt{|\tl g|} \left[ \sfrac{\m^2}{2} \tl{\R}  - \sfrac{1}{2} \tl g^{\mu \nu}\,\tl{\nabla}_{\!\mu}\,\varphi \, \tl{\nabla}_{\!\nu}\,\varphi - U(\varphi) \right]
\end{equation}
where the canonically normalized inflaton field $\varphi$ is related to the original nonminimally coupled scalar field $\phi$ by the expression
\begin{align}
    \left( \der{\varphi}{\phi}\right)^{\!\!2} = \sfrac{\m^2}{2F(\phi)} + \sfrac{3}{2} \m^2 \left( \sfrac{F'(\phi)}{F(\phi)}\right)^{\!\!2}
\end{align}
and the relation between potentials in two frames is
\begin{align}
    U[\varphi(\phi)] = \sfrac{V(\phi)}{\Omega^4(\phi)} \:.
\end{align}
Then, in flat FLRW spacetime defined in Eq.\ \autoref{eq:metric} Friedmann equation and the equation of motion for the scalar field in the EF become
\begin{subequations}
\begin{align}
    &\tl{H}^2 = \sfrac{1}{3 \m^2} \left[ \sfrac{1}{2} \dot{\varphi}^2 + U(\varphi)\right] \label{eq:EF_Friedmann} \\[2mm]
    &\ddot{\varphi} + 3 \tl{H} \dot{\varphi} + U'(\varphi) = 0 \label{eq:EF_scalar}
\end{align}
\label{eq:EF_equations}%
\end{subequations}
where, this time, an overdot and a prime represent the derivative with respect to transformed time variable $\tl{t}$ and the scalar field in the EF $\varphi$, respectively.

\section{THE VIABILITY OF SLOW-ROLL APPROACH\label{sec:viability_SR}}

\begin{figure}[h!]
	\centering
	\vspace{2mm}
	\includegraphics[width=0.52\textwidth]{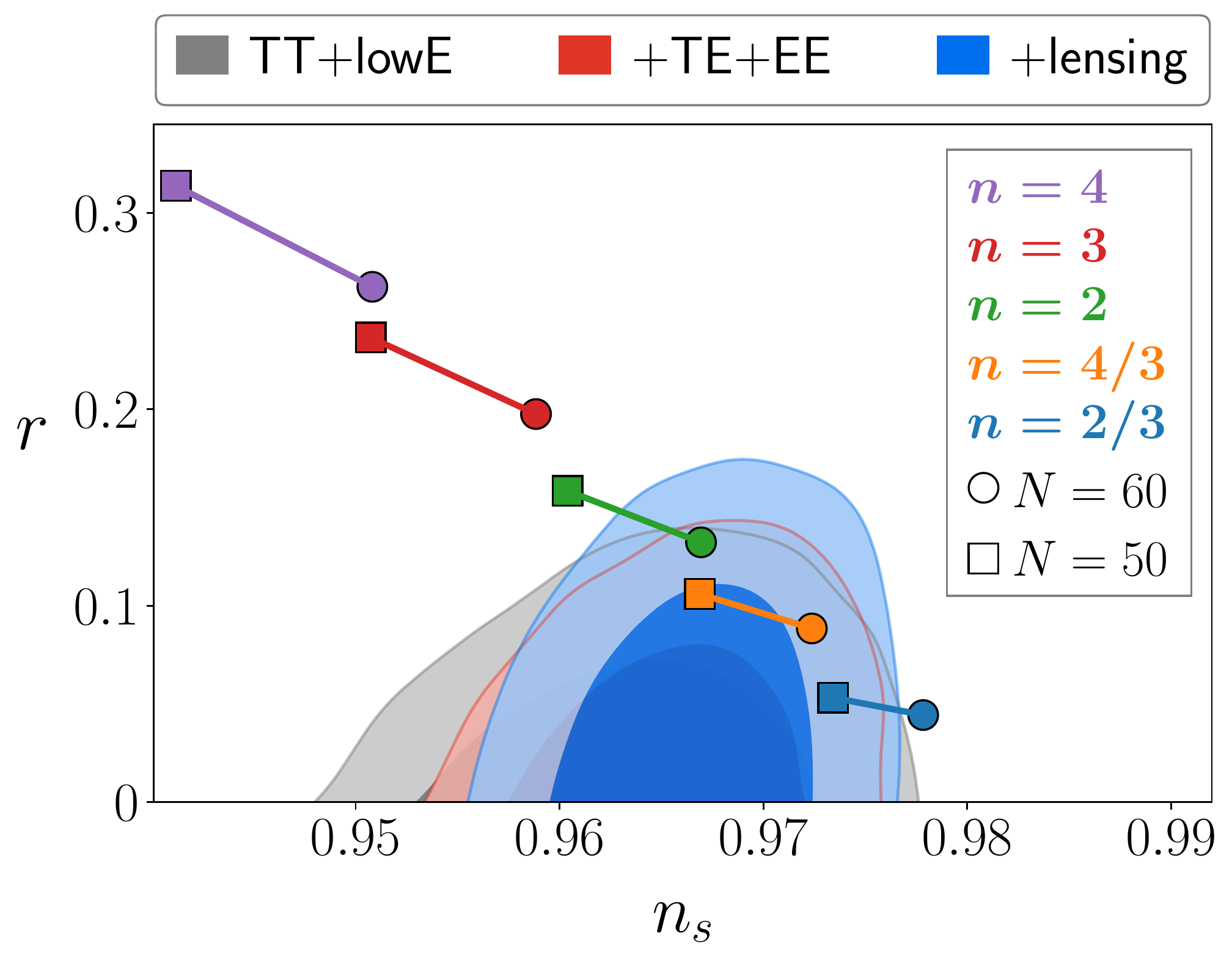}
	\vspace{-3mm}
	\caption{Inflationary observables $n_s$ and $r$ for inflaton field with monomial potential $V(\phi) \propto \phi^n$ with two different $e$-fold numbers, $N=50$ and $N=60$.}
\label{fig:inf_obs_minimal_monomial}
\end{figure}

Prior to discussion on the viability of SR approach in scalar-tensor theories it is appropriate to highlight some key points on minimally coupled scalar field models based on the recent observations. In order to test the predictions of a model by means of the observational data one way is to calculate the inflationary observables, $n_s$ and $r$, in terms of the SR parameters that are computed by applying the SR approximations to the equations of motion. As an example, for the minimally coupled scalar field model, which is obtained by setting $F(\phi)=\m^2/2$ in Eq.\ \autoref{eq:Friedmann_scalar_eqs}, if the potential of the scalar field is in the form of $V(\phi) \propto \phi^n$, $n_s-r$ graphs predicted by the model are obtained as shown in Fig.\ \autoref{fig:inf_obs_minimal_monomial} for some mostly used $n$ parameter values in the literature.

As seen from the figure, it is clear that the minimally coupled scalar field models with the monomial potential is barely compatible with the observational data for certain values of the potential parameter $n$. Therefore, in addition to the motivations coming from fundamental theories as indicated in Sec.\ \autoref{sec:intro}, the observational data also gives a hint to consider the alternatives such as scalar-tensor theories.

The SR approximation for any model has to be justified by the existence of inflationary attractors in the phase space of the corresponding dynamical system. When it comes to scalar-tensor theories, this issue requires an extra attention since there exist additional free parameters which can cause the phase space to be disrupted. This means that not every solution does follow the same inflationary pattern. Hence, it must be emphasized that without the proper inflationary attractors, all the scenarios following from this approach are far from being credible. In Ref.\ \cite{faraoni-gsr} this point is elaborated and a condition is given for the existence of the inflationary attractors in a specific model. Here a brief explanation is provided on the matter by the help of the dynamical system analysis methods.

To construct an autonomous dynamical system from Eq.\ \autoref{eq:Friedmann_scalar_eqs} one can choose two independent variables among $H$, $\phi$, $\dot{\phi}$, so that the dimension of the phase space is reduced. Then, in order to analyze the behaviour of the system one can find the fixed points and determine their characters which in turn specify the structure of inflationary attractors. To this end, here $H$ is eliminated from the equations and, with the definition $\dot{\phi} \equiv \psi$, the equation of motion for the scalar field turns into the following form
\begin{equation}
\left\{
\begin{array}{ll}
    \dot{\phi} = \psi \\[1ex]
    \dot{\psi} = -3H \psi + 6 \big(2H^2 + \dot{H}\big) F'(\phi) - V'(\phi)
\end{array}
\right.
\label{eq:dyn_sys}
\end{equation}
where
\begin{subequations}
\begin{align}
    H^2 &= \sfrac{1}{6 F(\phi)} \bigg[ \sfrac{1}{2} \psi^2 + V(\phi) - 6H \psi F'(\phi) \bigg] \label{eq:hubble_constraint} \\[2mm]
    \dot{H} &= \sfrac{-1}{2 F(\phi)} \bigg[ \bigg( \sfrac{1}{2} + F''(\phi) \bigg) \psi^2 + \big( \dot{\psi} - H \psi \big) F'(\phi) \bigg]
    \label{eq:drv_hubble_constraint} \:.
\end{align}
\end{subequations}
Then, the fixed points of Eq.\ \autoref{eq:dyn_sys} are obtained as
\begin{equation}
    \psi_\star = 0 \:, \qquad\quad 2\,\sfrac{F'(\phi_\star) \, V(\phi_\star)}{F(\phi_\star)} - V'(\phi_\star) = 0 \;.
\label{eq:fixed_points}
\end{equation}
It is obvious that the position of the fixed points lie on the $\psi=0$ axis in a model-independent way. On the other hand, unlike minimally coupled case, there exist invariant manifolds passing through these fixed points which separate the phase space and, therefore, cause to form different types of solutions. Consequently, this fact shows that not every initial condition describe a proper inflationary solution which can be defined as converging to an inflationary attractor providing the necessary amount of $e$-fold number, i.e. $50 \lesssim N \lesssim 60$, and converging to a vanishing scalar field that set the stage for the standard big bang cosmology. For instance, monomial potentials guarantee that one of the fixed points is $(\phi_\star,\psi_\star)=(0,0)$. If this point is stable and some solution with the proper initial conditions exists in the basin of attraction of this fixed point, the scalar field asymptotically vanishes. Then, one has to check that whether those initial conditions satisfy the condition on the amount of $e$-fold number, that is basically determined by the distance between the fixed points. Since the position of the fixed points depends on the structure of the coupling function, and subsequently the coupling parameter at hand, as seen in Eq.\ \autoref{eq:fixed_points}, it plays a major role to find the appropriate inflationary attractors and to apply the SR approximation to the system in a viable way.

\section{SLOW-ROLL EQUATIONS IN THE JORDAN FRAME \label{sec:SR_Eqs}}

The SR field equations in the JF are obtained in two ways : First the so-called GSR approximations \cite{torres97,morris01,faraoni-gsr} are applied to the system to get the approximate field equations assuming the existence of inflationary attractors \cite{faraoni-gsr} in phase space. Second after the SR field equations are written in the EF, they are expressed in terms of the JF variables by applying the conformal transformations defined in Eq.\ \autoref{eq:CT}, together with similar conditions to the GSR ones.

The SR parameters in the EF are defined as usual
\begin{equation}
    \SReps \equiv \sfrac{\m^{2}}{2 }\left[ \sfrac{U'(\varphi)}{U(\varphi)} \right]^{2} \;, \qquad
    \SReta \equiv  \m^{2}\left[ \sfrac{U''(\varphi)}{U(\varphi)} \right] \;, \qquad
    \zeta \equiv  \m^{2}\left[  \sfrac{U'(\varphi) \, U'''(\varphi)}{U^{2}(\varphi)} \right]^{1/2} .
\label{eq:SR_params_EF}
\end{equation}
To proceed in the EF one has to write $U$ in terms of the EF scalar field $\varphi$. However, since in general it is difficult to find $\varphi$ in terms of $\phi$ in closed form, the generally preferred strategy is to express each quantity of interest in terms of the JF quantities. The SR parameters, for example, are to be evaluated at $\varphi_{\text{hc}}$ which is the value of $\varphi$ at which the scales of interest cross the horizon during the inflationary epoch. Although calculation of the field value at horizon-crossing is not an easy task in both frames, by assuming that the scales of interest cross the horizon after $N$ $e$-folding before the end of inflation, we can write
\begin{equation}
    e^{N} \equiv \sfrac{\tl{a}(\tl{t}_{\rm end})}{\tl{a}(\tl{t}_{\rm hc})} = \sfrac{F_{\rm end}}{F_{\rm hc}} \sfrac{a(t_{\rm end})}{a(t_{\rm hc})} \:,
\label{eq:scales-comp}
\end{equation}
where $\phi_{\text{hc}}$ appearing in $F$ is the value of the JF scalar corresponding to $\varphi_{\text{hc}}$. This allows us to consider the SR parameters, mapped back to the JF, at correct time. Therefore we need SR field equations in the JF and need to solve them to get $a(t)$ and $\phi(t)$.

In the following subsections the SR field equations are obtained for both methods and calculation of the inflationary parameters are compared. Additionally, an approximation containing a higher-order term is also given to control the results.

\subsection{SR Equations in the JF via the GSR Conditions \label{ssec:generalized_SR}}

The dynamics of inflationary models with a single minimally coupled inflaton is considered in the ``SR approximation'' \cite{sr94} which amounts to the assumptions that the inflaton evolves slowly in comparison to the Hubble rate, and that the kinetic energy of the inflaton is smaller than its potential energy.  These conditions are expressed in a compact way as $|\ddot{\phi}| \ll H |\dot{\phi}| \ll H^{2}|\phi|$ and $\dot{\phi}^{2}\ll |V(\phi)|$. The generalization of these conditions to scalar-tensor theories with a coupling function $F(\phi)$, that has a sufficiently fast convergent Taylor expansion, is 
\begin{equation}
    |\ddot{F}| \ll H |\dot{F}|\ll H^{2}|F|
\label{eq:gsr}
\end{equation}
that was first pointed out in Ref.\ \cite{torres97}. Direct application of these conditions to the field equations in the JF, i.e.\ Eq.\ \autoref{eq:Friedmann_scalar_eqs}, leads to the following approximate forms
\begin{subequations}
\begin{align}
    H^{2} & \simeq \: \sfrac{V(\phi)}{6F(\phi)} \:, \label{JF-appfriedmann1} \\[2mm]
    3H\dot{\phi} & \simeq 2V(\phi)\,\sfrac{F'(\phi)}{F(\phi)} - {V'(\phi)} \label{JF-appscalar1}
\end{align}
\label{eq:SR_eom_generalized}%
\end{subequations}
which is the equation set used to calculate the SR conditions and inflationary variables in the JF.

\subsection{SR Equations in the JF via those of the EF \label{ssec:transformed_SR}} 

The SR field equations in the EF obtained from Eq.\ \autoref{eq:EF_equations} are
\begin{subequations}
\begin{align}
    \tl{H}^{2} \simeq & \: \sfrac{1}{3 \m^{2}} \, U(\varphi) \:, \label{EF-appfriedmann} \\[2mm]
    3\tl{H} \dot{\varphi} \simeq & - U'(\varphi)  \label{EF-appscalar} 
\end{align}
\end{subequations}
applying the SR conditions given in Eq.\ \autoref{eq:SR_params_EF}. The conformal transformation of these expressions in connection with the GSR conditions (Eq.\ \autoref{eq:gsr}) together, the SR field equations in the JF become
\begin{subequations}
\begin{align}
    H^{2} & \simeq \: \frac{1}{6F(\phi)} V(\phi) \:, \label{JF-appfriedmann2} \\[2mm]
    3H \dot{\phi} \left( 1 + 3\,\sfrac{[F'(\phi)]^2}{F(\phi)} \right) & \simeq 2V(\phi)\, \sfrac{F'(\phi)}{F(\phi)} - V'(\phi) \:. \label{eq:JF-appscalar2}
\end{align}
\label{eq:SR_eom_EF_to_JF}%
\end{subequations}
The SR approximated Friedmann equation, i.e.\ Eq.\ \autoref{JF-appfriedmann2}, is exactly the same as the one obtained in the previous section but the SR approximated scalar field equation is different from Eq.\ \autoref{JF-appscalar1} derived in the GSR method. 

The difference between the scalar field equations in the JF implies that the results of calculation of $\phi_{\rm hc}$ are different, and thus $\varphi_{\rm hc}$ and SR parameters are different in turn.  As the data becomes sharpen, this difference may lead to important difference between the observed and theoretically calculated values of the inflationary parameters.

\subsection{Comparison of SR Parameters and Inflationary Observables
\label{ssec:FrameComparison}}

Here the SR parameters and the inflationary observables, $n_s$ and $r$, are calculated using the results obtained by two different approaches given above. To begin with, the following SR parameters
\begin{equation}
    \SReps_H \equiv - \sfrac{\dot{H}}{H^2} \:, \qquad \SReta_H \equiv \sfrac{\dot{\SReps}_H}{H \SReps_H} \:; \qquad \SReps_F \equiv \sfrac{\dot{F}}{HF} \:, \qquad \SReta_F \equiv \sfrac{\dot{\SReps}_F}{H \SReps_F}
\label{eq:SR_param}
\end{equation}
are defined and the inflationary variables in terms of these SR parameters
\begin{subequations}
\begin{align}
    r &= 8\,(2\SReps_H + \SReps_F) \:, \\[2mm]
    n_s &= 1 - 2 \SReps_H - \SReps_F - \sfrac{2 \SReps_H\,\SReta_H + \SReps_F\,\SReta_F}{2 \SReps_H + \SReps_F}
\end{align}
\label{eq:inf_obs}%
\end{subequations} 
given in Ref.\ \cite{granda2019} are taken into account. Then, the equations of motion in Eq.\ \autoref{eq:SR_eom_generalized}, which are obtained by applying the GSR conditions, yield the SR parameters as
\begin{equation}
\begin{aligned}
    \SReps_H &= F \left( \sfrac{F'}{F} - \sfrac{V'}{V} \right)\!\!\left( 2\sfrac{F'}{F} - \sfrac{V'}{V} \right) \:, \qquad
    \SReta_H = 2 \left( \sfrac{F'}{F} - \sfrac{V'}{V} \right)^{-1} \SReps'_H \:, \\[2mm]
    \SReps_F &= 2F' \left( 2 \sfrac{F'}{F} - \sfrac{V'}{V} \right) \:, \qquad
    \SReta_F = 8F'' - 2F \left[ \sfrac{\big(F'V'\big)'}{F'V} + 2\left(\sfrac{F'}{F}\right)^{\!\!2} - \left(\sfrac{V'}{V}\right)^{\!\!2}\right]
\end{aligned}
\label{eq:SR_param_JF}
\end{equation}
in terms of the coupling function and the potential of the scalar field.

On the other hand, if these SR parameters are calculated by Eq.\ \autoref{eq:SR_eom_EF_to_JF}, i.e.\ the equations transformed from EF to JF, and the results are compared with the ones obtained by the first method, the relation between SR parameters and, consequently, the inflationary observables computed by two method is found to be
\begin{equation}
    A_{(1)} = A_{(2)} \left[ 1 + 3 \sfrac{(F')^2}{F} \right] \;, \qquad A = \SReps_{H/F},\,\SReta_{H/F},\,r,\,n_s-1
\label{eq:params_difference}
\end{equation}
where the subscripts (1) and (2) represent the parameters obtained by the GSR approach in the JF and the ones transformed from EF to JF, respectively.

Additionally, the $e$-fold integrals are defined as
\begin{equation}
    N_{(1)} = \int\limits_{\phi_{\rm i}}^{\phi_{\rm e}} \sfrac{H}{\dot{\phi}}\,\d \phi = \sfrac{1}{2} \int\limits_{\phi_{\rm i}}^{\phi_{\rm e}} \sfrac{V}{2 V F' - V' F} \,\d \phi \:.
\label{eq:N_JF}
\end{equation}
and 
\begin{equation}
    N_{(2)} = \int\limits_{\varphi_{\rm i}}^{\varphi_{\rm e}} \sfrac{\tilde{H}}{\dot{\varphi}}\,\d \varphi = \sfrac{1}{2} \int\limits_{\phi_{\rm i}}^{\phi_{\rm e}} \left[ 1 + 3 \sfrac{(F')^2}{F} \right] \sfrac{V}{2 V F' - V' F} \,\d \phi \:.
\label{eq:N_EF_to_JF}
\end{equation}
for two methods.

\subsection{SR Equations in the JF via the Higher Order SR Conditions \label{ssec:HSR}}
The SR approximation has to be applied meticulously for the non-minimal coupling case since the GSR approach \cite{torres97,morris01,granda2019} might be imprecise in the sense of preserving attractor structure as illustrated in Sec.\ \autoref{sec:viability_SR}. Thus, comparison of the above analysis, i.e. GSR and EF-to-JF methods, with a stricter one may be instructive. Here, without using any conformal transformation tool the analysis in the JF is performed by keeping a higher order term ($\dot{H}$) in the equation of motion. Following this manner, Eqs.\ \autoref{eq:JF_Friedmann} and \autoref{eq:JF_Scalar} become
\begin{subequations}
    \label{eq:JF_app_eqs}
    \begin{align}
        H^2 &\simeq \frac{V}{3F} \label{eq:JF_app_Friedmann}\\[1ex]
        3H\dot{\phi} &\simeq 6H^2 F' + 3\dot{H}F' - V' \label{eq:JF_app_EoM}
    \end{align}    
\end{subequations}
where $\dot{H}$, which differs from the original one\footnote{Although it is possible to include approximated $\dot{H}$ expression by applying the SR conditions of this approach to Eq.\ \autoref{eq:JF_Acceleration} directly, reproducing $\dot{H}$ from Eq.\ \autoref{eq:JF_app_Friedmann} is preferred due to computational convenience and, more importantly, comparability of analytical representations with the results of previously given methods.}, can be obtained from Eq. \autoref{eq:JF_app_Friedmann} as
\begin{align}
    \label{eq:dotH_different}
    \dot{H} = \sfrac{\dot{\phi}}{6HF^2}(V'F - VF') = \sfrac{\dot{\phi}}{6H} \left(\sfrac{V}{F}\right)'
\end{align}
together with the following expression 
\begin{align}
    \label{eq:pdoh}
    \sfrac{\dot{\phi}}{H} = \sfrac{F(4V'F' - V'F)}{3 V (F')^2 + 2F} = \left[ 2 + 3 \sfrac{(F')^2}{F} \right]^{-1} F V' \left( 4\sfrac{F'}{F} - 1 \right) \;.
\end{align}
Using the set given by Eq. \autoref{eq:JF_app_eqs}, the SR parameters can be written as
\begin{align}
    \begin{aligned}
        \SReps_H &= \left[ 2 + 3 \sfrac{(F')^2}{F} \right]^{-1} F \left( \sfrac{F'}{F} - \sfrac{V'}{V} \right)\!\!\left( 2\sfrac{F'}{F} - \sfrac{V'}{V} \right) \; &&, \quad \SReta_H = \sfrac{\dot{\phi}}{H} \sfrac{\SReps'_H}{\SReps_H} \\[1ex] 
        \SReps_F &=  \left[ 2 + 3 \sfrac{(F')^2}{F} \right]^{-1} 2F' \left( 2 \sfrac{F'}{F} - \sfrac{V'}{V} \right) \;&&, \quad \SReta_F = \sfrac{\dot{\phi}}{H} \sfrac{\SReps'_F}{\SReps_F}\\[1ex]
    \end{aligned}
\end{align}
where $\dot{\phi}/H$ is given by Eq. \autoref{eq:pdoh}
and, furthermore, number of $e$-folding integral is
\begin{align}
    N = \int H \d t = \int\limits_{\phi_i}^{\phi_e} \sfrac{H}{\dot{\phi}}\,\d \phi = \int\limits_{\phi_i}^{\phi_e} \sfrac{3 V F'^2 + 2F}{F(4V'F' - V'F)}\, \d \phi \;.
\end{align}

\clearpage

\section{AN EXAMPLE : NONMINIMALLY COUPLED SCALAR FIELD \label{sec:example}}

\begin{figure*}[!t]
	\centering
	\hspace{-5mm}
	\begin{tabular}{@{}c@{}}
		\includegraphics[width=.52\linewidth]{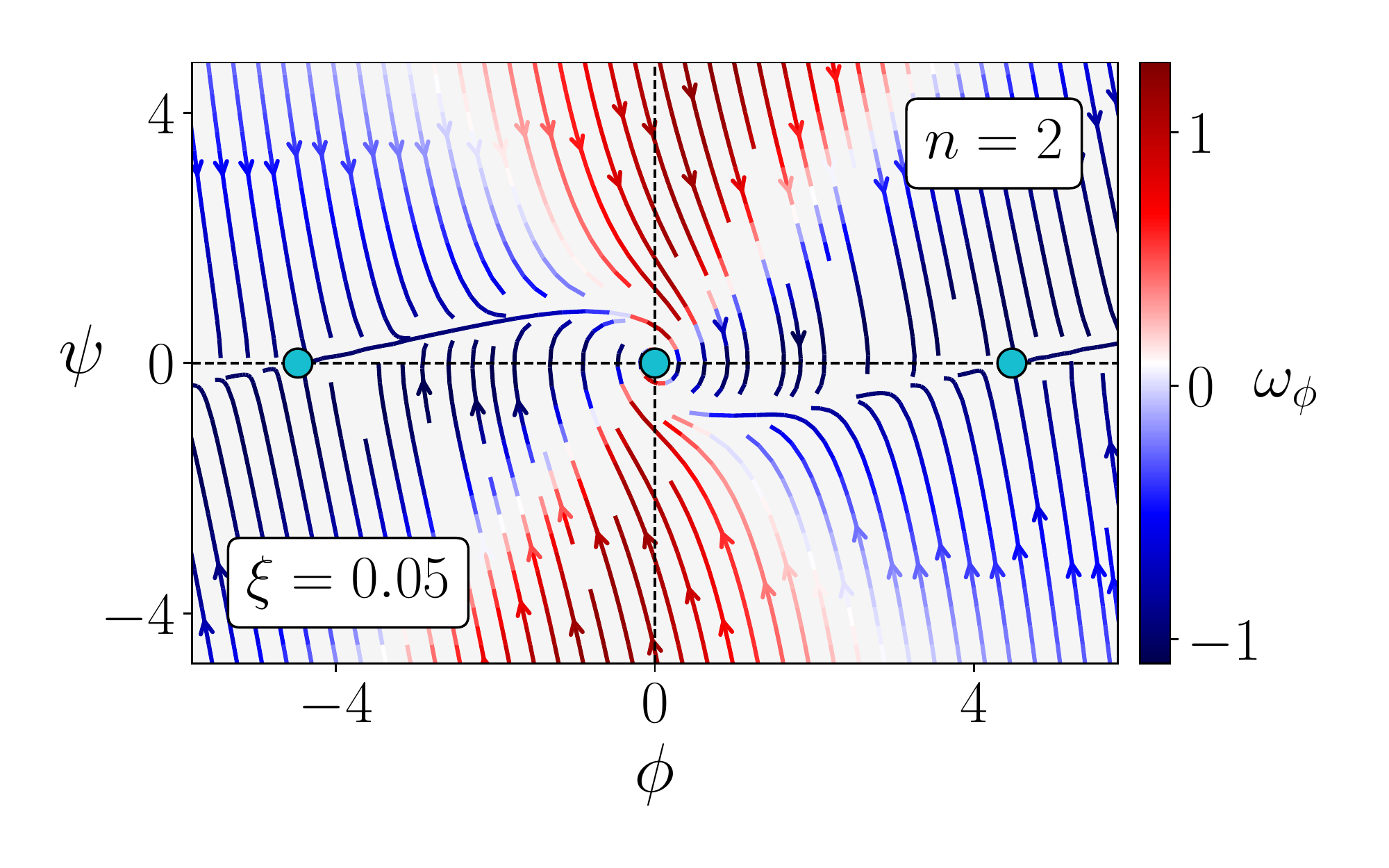} \\
	\end{tabular} \hspace{-6mm}
	\begin{tabular}{@{}c@{}}
		\includegraphics[width=.52\linewidth]{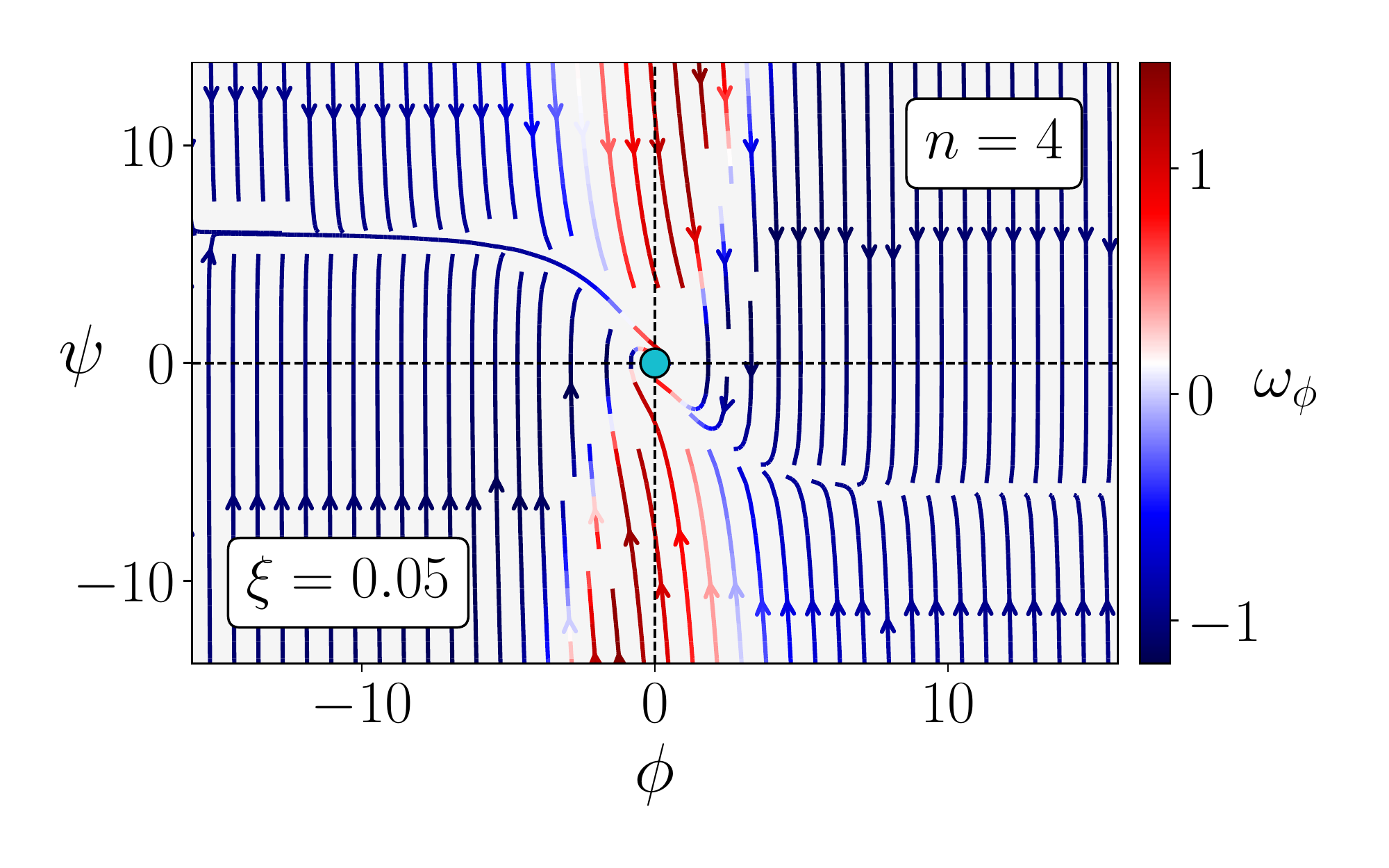} \\
	\end{tabular}
	
	\vspace{-5mm}
	
	\hspace{-5mm}
	\begin{tabular}{@{}c@{}}
		\includegraphics[width=.52\linewidth]{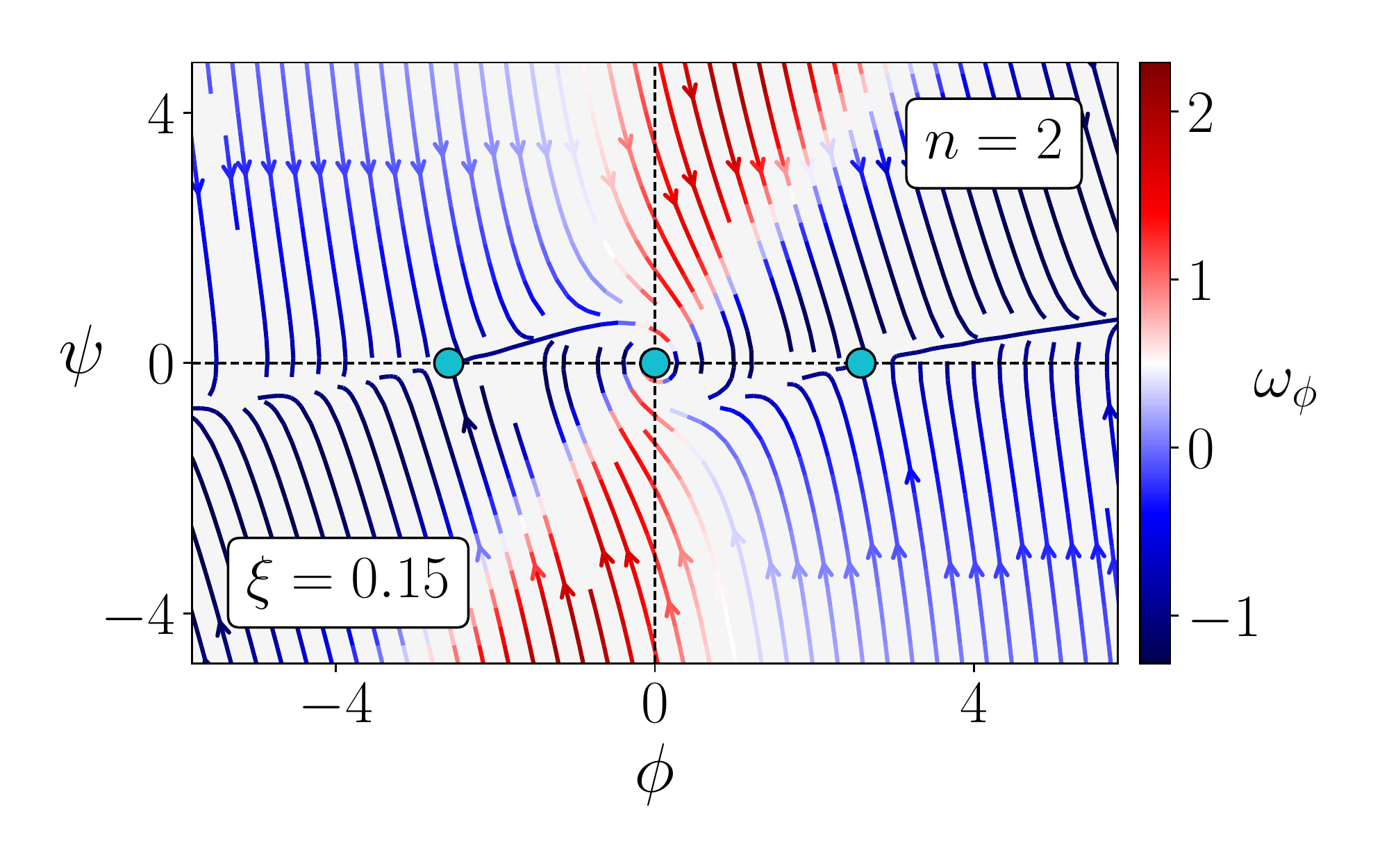} \\
	\end{tabular} \hspace{-6mm}
	\begin{tabular}{@{}c@{}}
		\includegraphics[width=.52\linewidth]{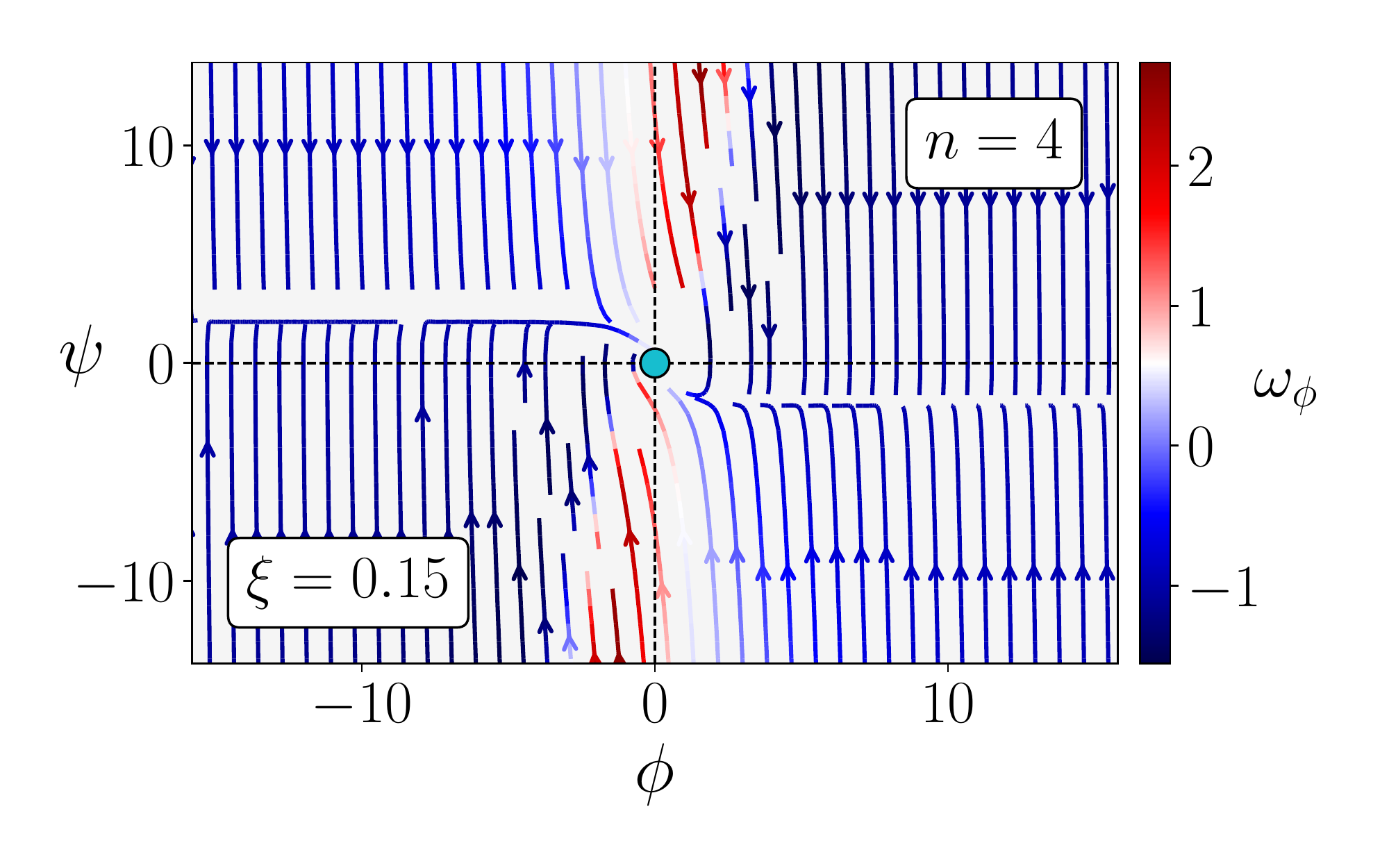} \\
	\end{tabular}
	
	\vspace{-6mm}
	
	\caption{The phase spaces of the system given in Eq.\ \autoref{eq:dyn_sys} together with Eq.\ \autoref{eq:coupling_potential}. Left (right) column corresponds to $n=2$ ($n=4$) for the potential with two different values of the coupling constant. Here $\m^2=1$ for simplicity. Dots in the phase planes are the fixed points of the system given in Eq.\ \autoref{eq:fixed_points}.}
\label{fig:non_minimal_phase}
\end{figure*}

In order to exemplify the claims mentioned in the previous two sections the coupling function and the potential of the scalar field are chosen in the following forms
\begin{equation}
    F(\phi) = \sfrac{1}{2} \left( \m^2 + \xi \phi^2 \right) \:, \qquad V(\phi) = V_{\!o}\,\phi^n \:.
\label{eq:coupling_potential}
\end{equation}
Then, the fixed points in Eq.\ \autoref{eq:fixed_points} become 
\begin{equation}
    \psi_\star = 0 \:; \quad\qquad \phi^{\,(1)}_\star = 0 \:, \qquad \phi^{\,(2,3)}_\star = \pm \sqrt{\sfrac{n\,\m^2}{\xi(4-n)}}
\end{equation}
where for $n=1$ and $n=4$ the fixed points $\phi^{(1)}_\star$ and $\phi^{(2,3)}_\star$, respectively, do not exist. A dynamical system analysis for this model was studied in Ref.\ \cite{Amendola1990} in detail and a condition for the stability of the solutions was provided in Ref.\ \cite{faraoni-gsr}. Therefore, those issues are not discussed further here, instead, to illustrate the aforementioned claims, the form of the solutions obtained by solving Eq.\ \autoref{eq:dyn_sys} numerically are represented in Fig.\ \autoref{fig:non_minimal_phase} for different values of the coupling parameter, $\xi$, and for the potentials with $n = 2$ and $n = 4$. It seems that $n=4$ is the exceptional case for this model and all solutions converge to the inflationary attractor before reaching the fixed point at the origin. Nevertheless, for the cases with $n\neq4$, it can be seen that as the value of the coupling constant increases, the fixed points get closer to each other and this makes the length of the appropriate inflationary attractors smaller in the region between the fixed points. As a result, the initial conditions that give rise to a solution, to which the SR approximation can be applied properly, are restricted by the value of the coupling constant in addition to the $e$-fold number. On the other hand, outside that region, i.e. outside the basin of attraction of the fixed point at the origin, the solutions are divergent although they obey $\omega_\phi\approx-1$.

To conclude this debate, it can be stated that the presence and the structure of the inflationary attractors severely depend on the value of the coupling parameter $\xi$ which in turn implies that blindly applying the SR approximation leads to wrong conclusions. Therefore, before applying the SR approximation one has to check the phase space structure and determine the range of free parameters in the model.

Following the dynamical system analysis, two previously explained methods are applied to the model to calculate the inflationary variables and to compare the predictions in the light of the observational data. For the model at hand the SR parameters in Eq.\ \autoref{eq:SR_param_JF} obtained by the GSR method become
\begin{equation}
\begin{aligned}
    \SReps_H &= \sfrac{1}{2} \left[ (n-2)(n-4)\xi + \m^2\,\sfrac{n^2\m^2 + (n^2-8)\xi\phi^2}{\phi^2(\m^2+\xi\phi^2)} \right] \;, \\[2mm]
    \SReta_H &= (4-n)\xi + \sfrac{n\m^2}{\phi^2} - \sfrac{(4-n)\xi\phi^2-n\m^2}{(2-n)\xi\phi^2-n\m^2}\bigg[ (n+2)\xi + \sfrac{n\m^2}{\phi^2} - \sfrac{4\xi^2\phi^2}{\m^2+\xi\phi^2} \bigg] \;, \\[2mm]
    \SReps_F &= 2\xi\, \sfrac{(4-n)\xi\phi^2-n\m^2}{\m^2+\xi\phi^2} \;, \\[2mm]
    \SReta_F &= \sfrac{8\xi\m^2}{\m^2+\xi\phi^2}
\end{aligned}
\label{eq:SR_params_nmc}
\end{equation}
and the exit value of the scalar field, which is determined by the solution of the equation stemming from the condition $\SReps_H(\phi_{\rm e})=1$, is obtained as
\begin{equation}
    \sfrac{\phi_{e}^2}{\m^2} = \sfrac{-1+n\xi(n-3) \pm \sqrt{(n\xi+1)^2 + 4n\xi}}{\xi\big[2-(n-2)(n-4)\xi\big]} \;.
\label{eq:phi_end_sq_JF}
\end{equation}
This expression naturally brings about following two conditions
\begin{equation}
    \phi_e^2 \geq 0 \;, \qquad (n\xi+1)^2 + 4n\xi \geq 0
\label{eq:phi_end_sq_JF_constraint}
\end{equation}
that constraint the values of the potential parameter $n$, and the coupling constant $\xi$. Furthermore, the initial value of the scalar field in terms of the exit value and $e$-fold number is obtained from the solution of Eq.\ \autoref{eq:N_JF} as
\begin{equation}
\phi_i^2 = \left\{
\begin{array}{ll}
    \phi_e^2 + 8\m^2 N &\;,\quad n = 4 \\[3mm]
    \left[ \phi_e^2 + \sfrac{n\m^2}{\xi(n-4)} \right] e^{-2\xi(n-4)N} - \sfrac{n\m^2}{\xi(n-4)} &\;,\quad n \neq 4 \\[3mm]
\end{array}
\right.
\label{eq:phi_initial_sq_JF}
\end{equation}
which also can be recast into the following form 
\begin{equation}
\phi_i^2 = \left\{
\begin{array}{ll}
    \phi_e^2 + 8\m^2 N &\;,\quad n = 4 \\[3mm]
    \big( \phi_e^2 - \phi_\star^2\,\big)\,e^{2nN\m^2/\phi_\star^2} + \phi_\star^2 &\;,\quad n \neq 4 \\[1mm]
\end{array}
\right.
\label{eq:phi_initial_sq_JF2}
\end{equation}
where $\phi_\star$ is the fixed point of the system. This shows that the position of the fixed point determines the proper initial conditions converging to the inflationary attractors. Therefore, in addition to $e$-fold number $N$, value of the coupling constant also effects the validity of SR approximation as mentioned in the previous section since value of the fixed point depends on the coupling constant.

On the other hand, in the second method, i.e.\ in the EF-to-JF transformed frame, the SR parameters are calculated by plugging the results given in Eq.\ \autoref{eq:SR_params_nmc} together with the coupling function into Eq.\ \autoref{eq:params_difference}. Then, following the same arguments above, the exit value of the scalar field is obtained as follows
\begin{equation}
    \sfrac{\phi_{e}^2}{\m^2} = \sfrac{-1+n\xi(n-3) \pm \sqrt{(3n\xi+1)^2 + (2n\xi)^2}}{\xi\big[2(1+6\xi)-(n-2)(n-4)\xi\big]}
\label{eq:phi_end_sq_TF}
\end{equation}
which, this time, yields only one mathematical constraint $\phi_e^2 \geq 0$. Additionally, the solution of Eq.\ \autoref{eq:N_EF_to_JF} yields the initial value of the scalar field as
\begin{equation}
\left\{
\begin{array}{ll}
    \big(\phi_i^2 - \phi_e^2\big)\!\left( 1 - \sfrac{3\xi}{4\m^2} \right) -  \sfrac{3}{4\m^2} \ln\!\left( \sfrac{\xi\phi_e^2 - 1}{\xi\phi_i^2 - 1} \right) = 8\m^2 N &\;,\quad n = 4 \\[5mm]
    \ln\!\left( \left[ \sfrac{\xi(n-4)\phi_i^2 + n\m^2}{\xi(n-4)\phi_e^2 + n\m^2} \right]^{1-3n/2\xi} \left[ \sfrac{\m^2+\xi\phi_i^2}{\m^2+\xi\phi_e^2} \right]^{3(n-4)/2\xi} \right) = 2\xi(4-n)N &\;,\quad n \neq 4 \\[3mm]
\end{array}
\right.
\label{eq:phi_initial_sq_TF}
\end{equation}

\begin{figure*}[!t]
	\centering \hspace{-7mm}
	\begin{tabular}{@{}c@{}}
		\includegraphics[width=.53\linewidth]{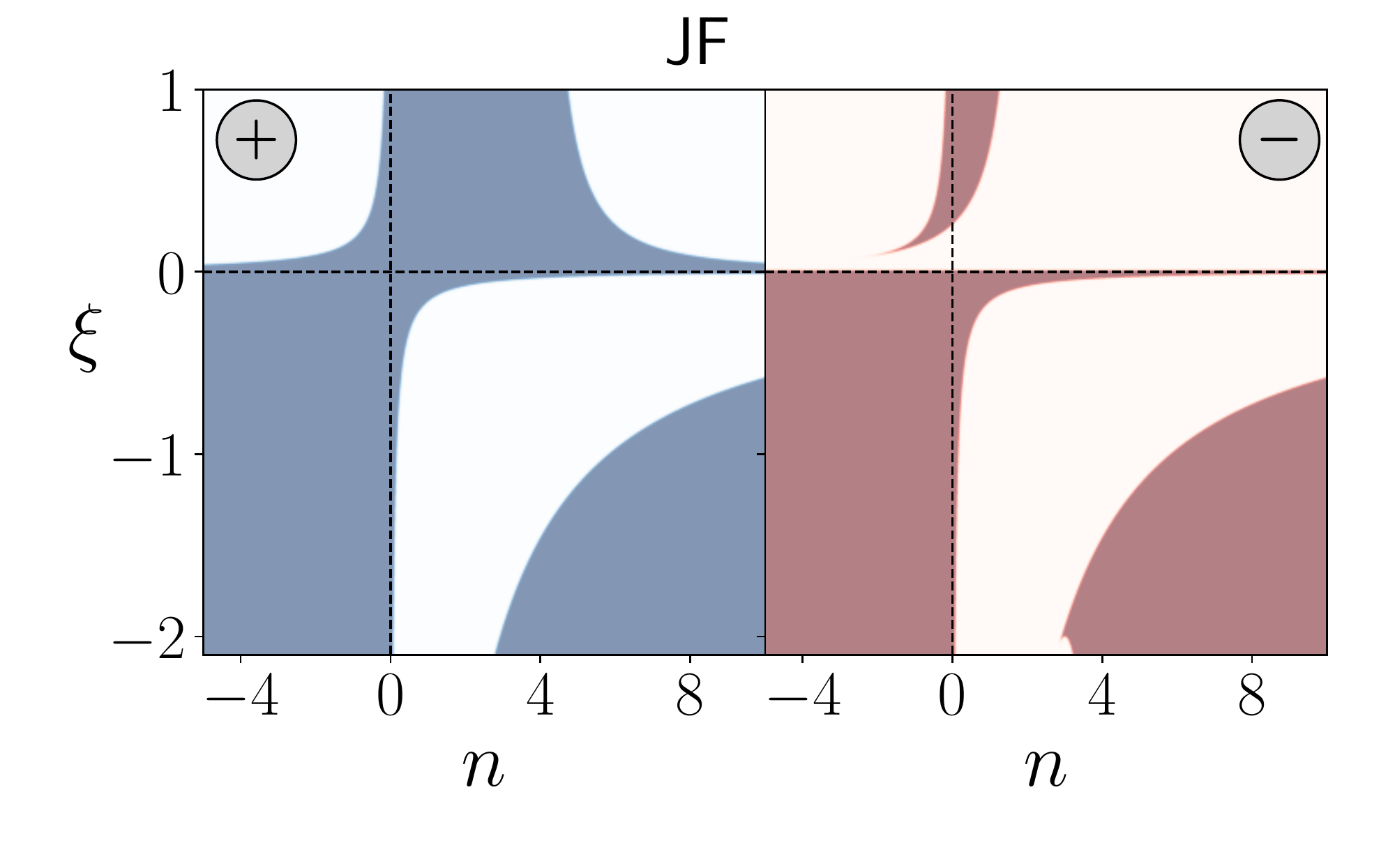} \hspace{-5mm} \\
	\end{tabular}
	\begin{tabular}{@{}c@{}}
		\includegraphics[width=.53\linewidth]{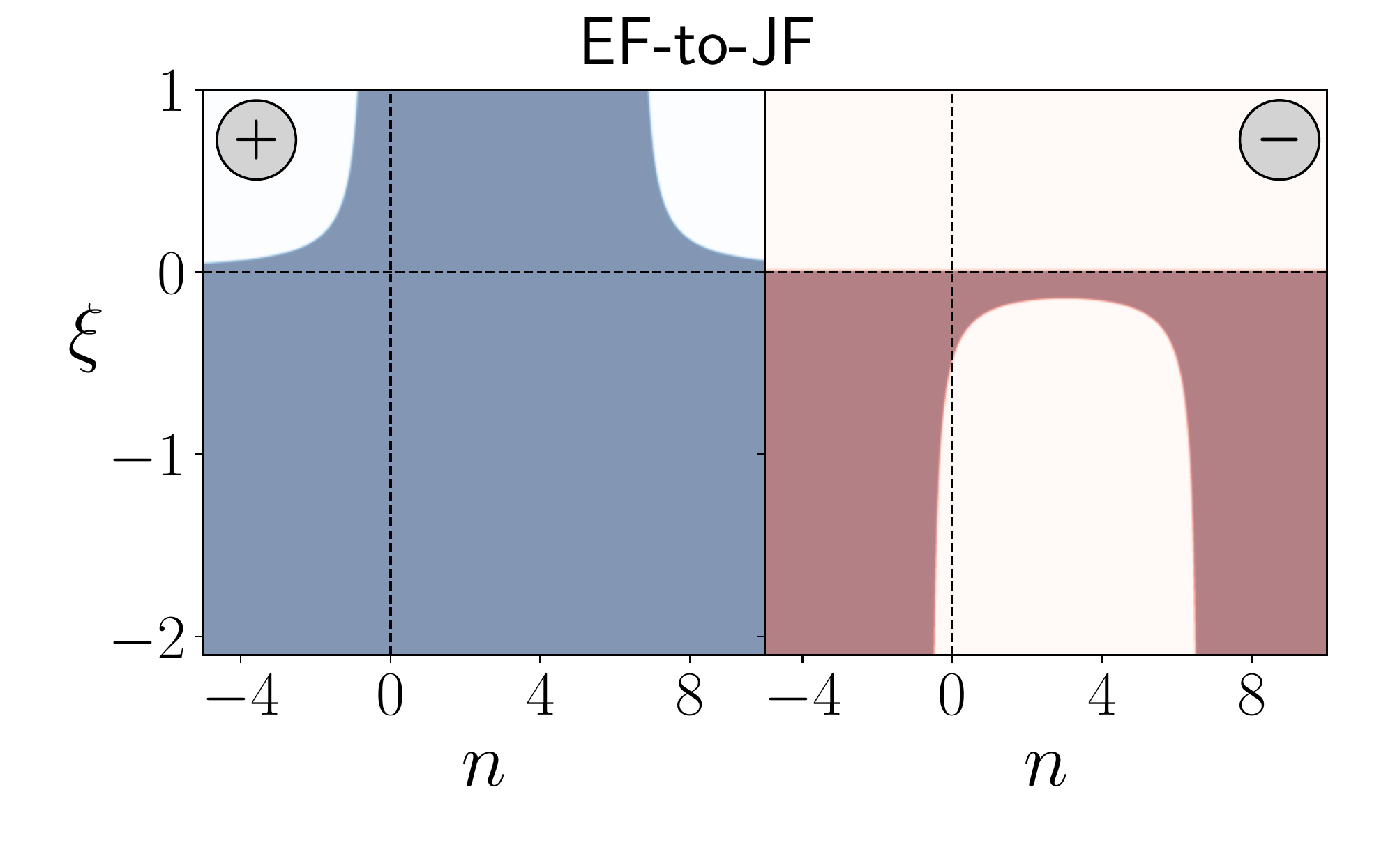} \hspace{-6mm} \\
	\end{tabular}

	\vspace{-7mm}
	
	\caption{Parameter spaces of the potential and the coupling parameters for the GSR (left) and the EF-to-JF (right) methods given by Eqs.\ \autoref{eq:phi_end_sq_JF} and \autoref{eq:phi_end_sq_TF}, respectively. Only the values in the shaded regions are allowed. Plus and minus signs in the plots correspond to the same signs in the roots of $\phi_e^2/\m^2$ for both methods.} \vspace{2mm}
\label{fig:parameter_spaces}
\end{figure*}

Before taking any further steps towards investigation of $n_s-r$ relations, it is useful to examine the outcomes of the inequalities between $n$ and $\xi$ coming from Eqs.\ \autoref{eq:phi_end_sq_JF} and \autoref{eq:phi_end_sq_TF}. For the one obtained in the GSR method the expression on the left in Eq.\ \autoref{eq:phi_end_sq_JF_constraint}, namely $\phi_e^2 \geq 0$, is more restrictive than the other one, and it is also the one and only constraint for the EF-to-JF method. Therefore, the condition $\phi_e^2 \geq 0$ is enough to examine the relation between the parameters of the model for both approaches. The graphical illustrations of these constraints are shown in Fig.\ \autoref{fig:parameter_spaces}. First thing to notice from the figure is that the roots with positive sign for both methods allow more values especially for $n>0$ and $\xi>0$ that describes the part of the parameter space which is primarily focused on in this work. In that region it seems that the increasing $n$ values naturally restrict the values of $\xi$. However, these conditions constrain strong couplings since the small values of $\xi$ are still applicable. Another point to mention is that for the region $n>0$ and $\xi<0$ the values in the GSR method is restricted whereas in the EF-to-JF method all values are acceptable.

\begin{figure*}[!t]
	\centering \hspace{-3mm}
	\begin{tabular}{@{}c@{}}
		\includegraphics[width=.56\linewidth]{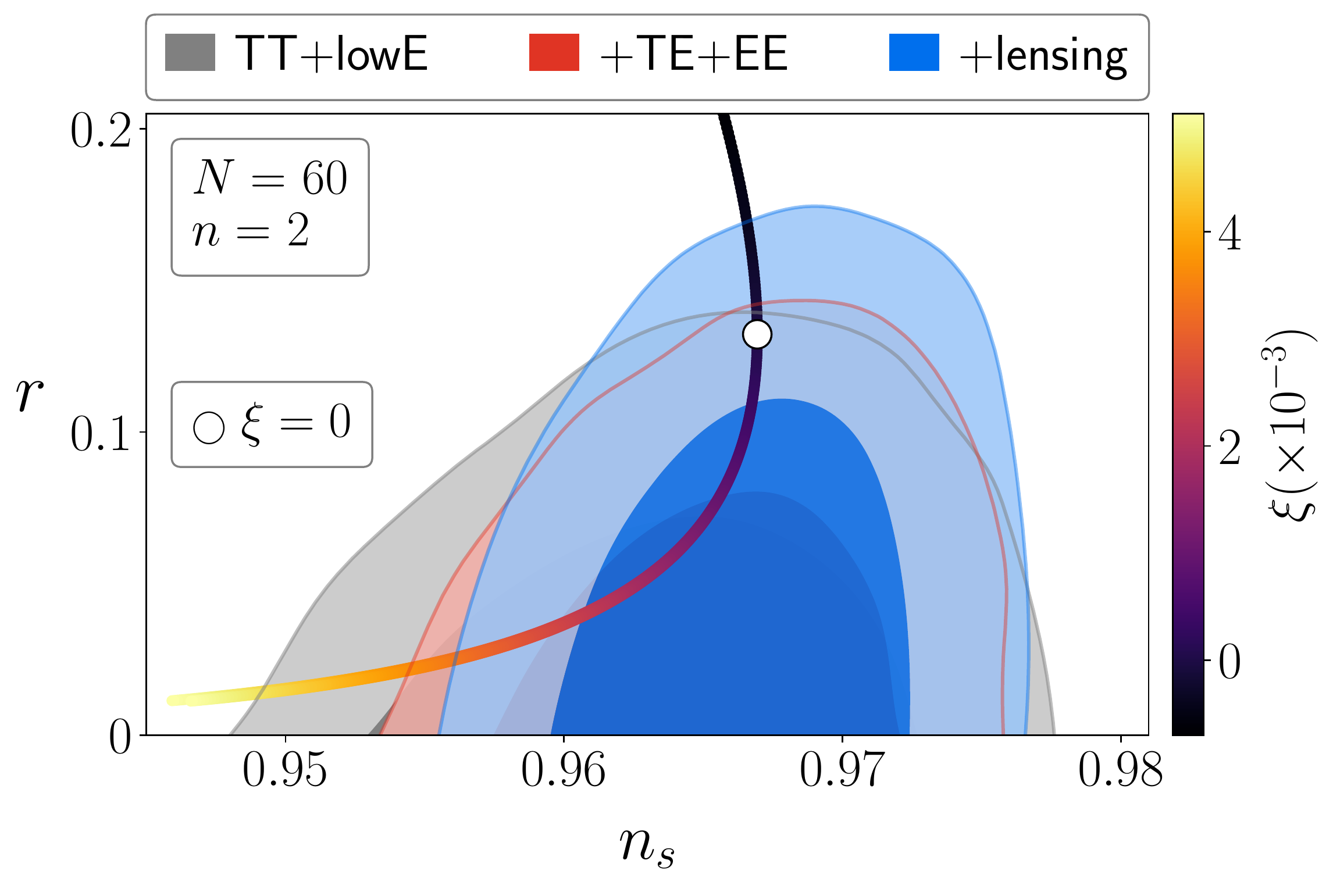} \hspace{0mm} \\
	\end{tabular}
	\begin{tabular}{@{}c@{}}
		\includegraphics[width=.42\linewidth]{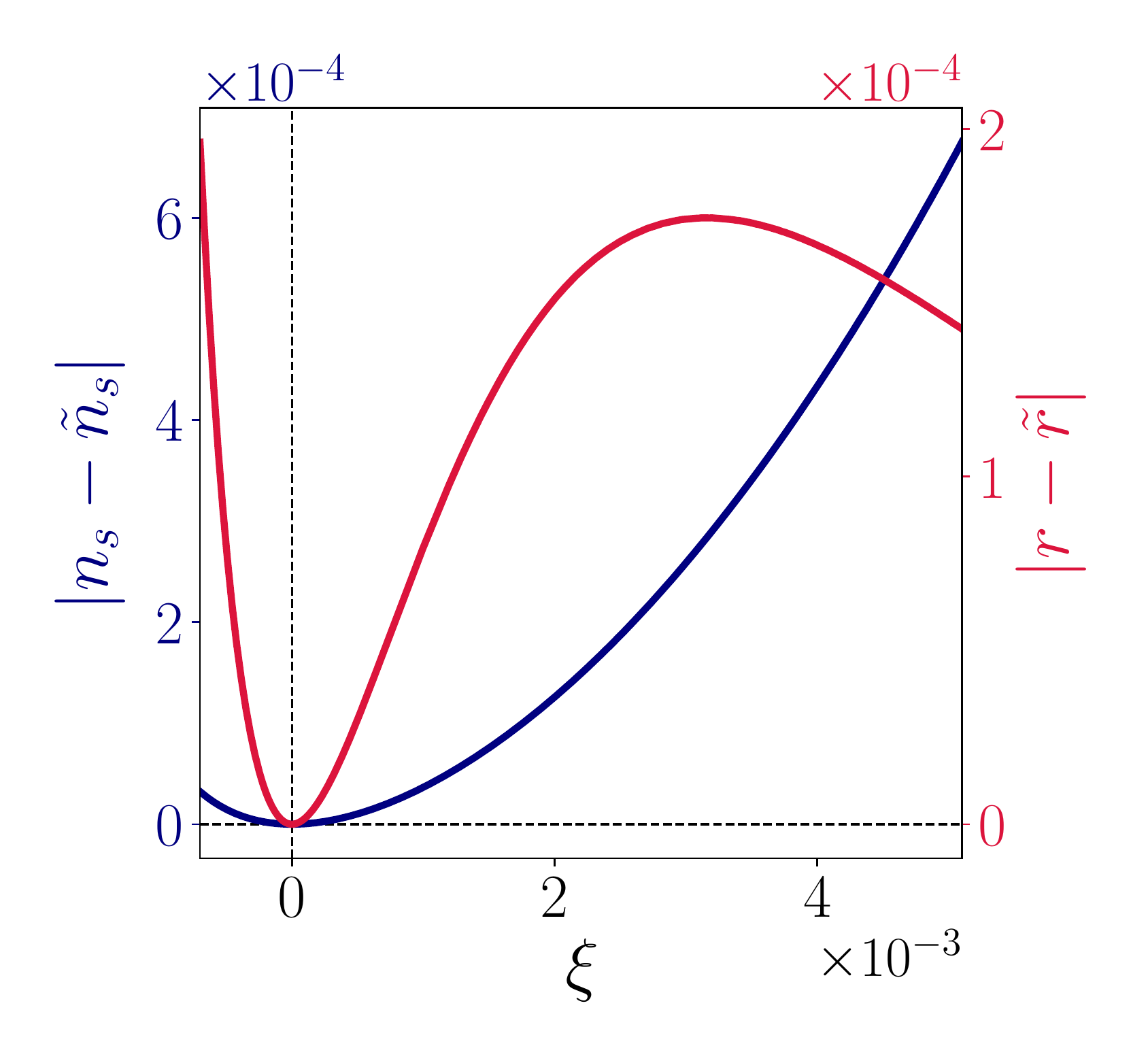} \\
	\end{tabular}
	
	\vspace{-2mm}
	
	\begin{tabular}{@{}c@{}} \hspace{-3mm}
		\includegraphics[width=.56\linewidth]{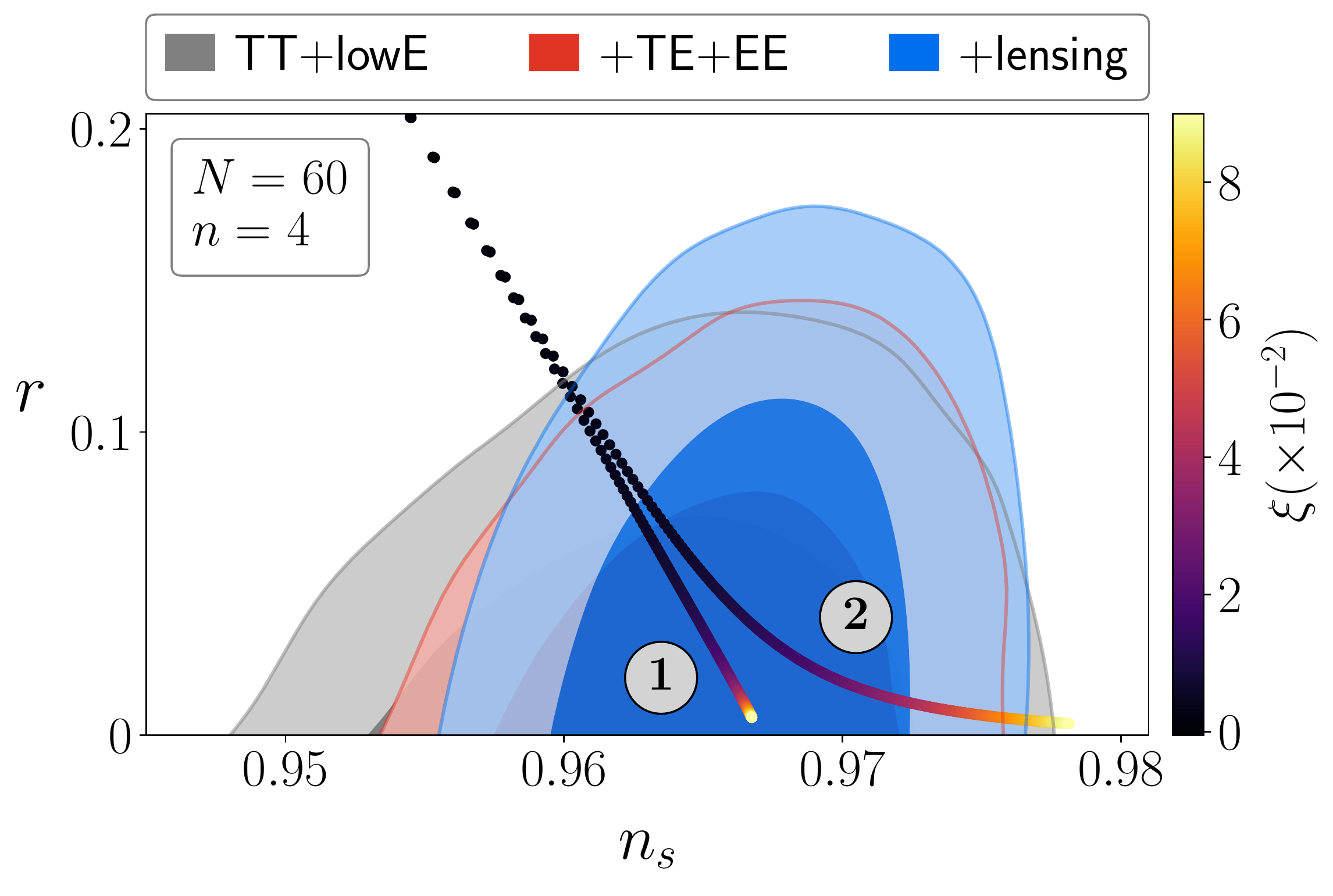} \hspace{0mm} \\
	\end{tabular}
	\begin{tabular}{@{}c@{}}
		\includegraphics[width=.42\linewidth]{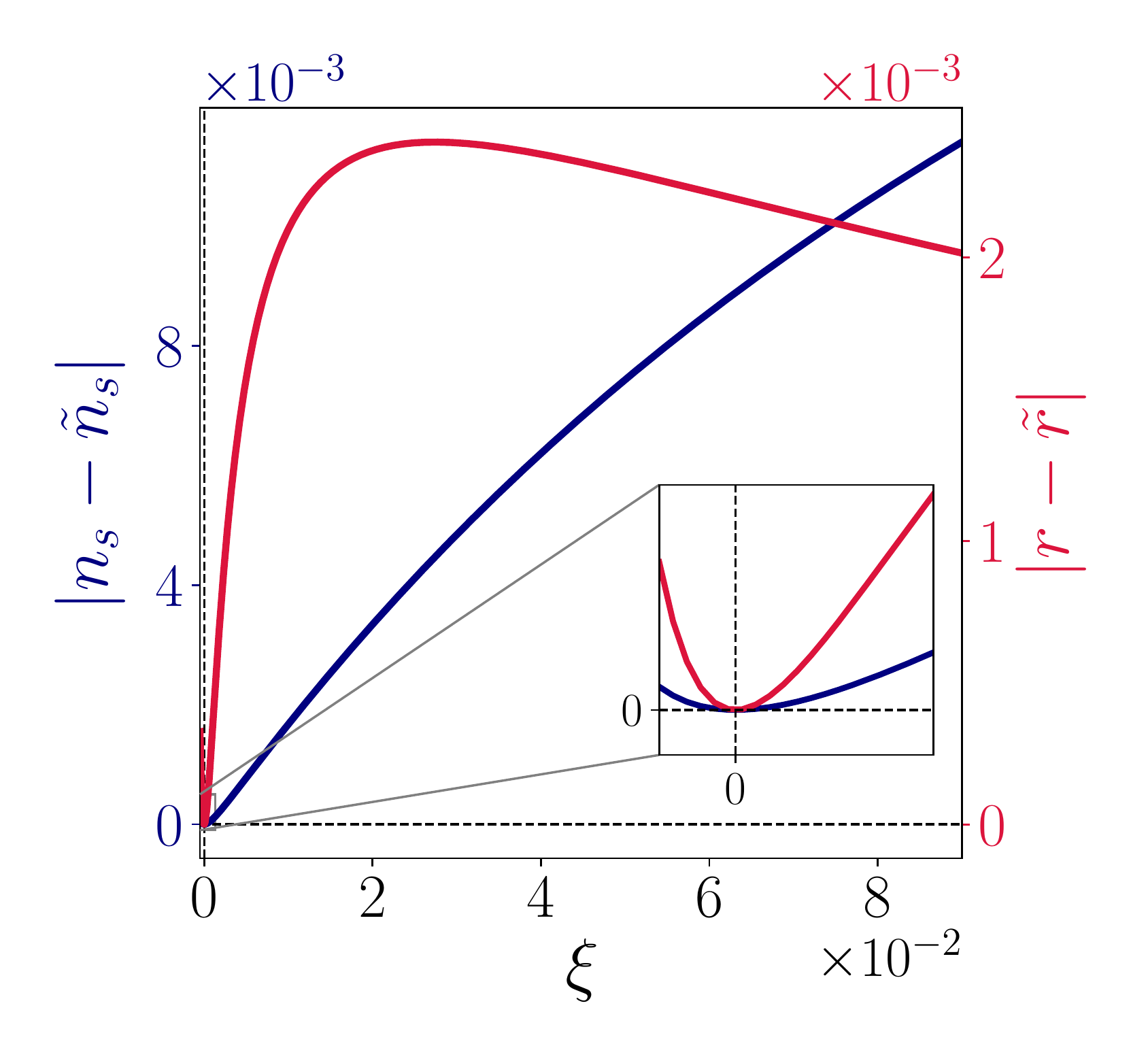} \\
	\end{tabular}
	
	\vspace{-5mm}
	
	\caption{$n_s-r$ graphs for the potential $V=V_{\!o}\,\phi^n$ with two parameters $n=2$ (top panel) and $n=4$ (bottom panel) with illustration of the difference of inflationary variables. Here tilde stands for the method of EF-to-JF. (1) and (2) in the bottom panel represent the solutions obtained by the GSR and EF-to-JF methods, respectively.}
\label{fig:ns_r_60}
\end{figure*}

In order to compute $n_s-r$ relations for the GSR method [the EF-to-JF method] one can plug Eq.\ \autoref{eq:phi_end_sq_JF} [Eq.\ \autoref{eq:phi_end_sq_TF}] into Eq.\ \autoref{eq:phi_initial_sq_JF} [Eq.\ \autoref{eq:phi_initial_sq_TF}] and then use the resulting expression for $\phi_i^2$ in Eq.\ \autoref{eq:SR_params_nmc} together with Eq.\ \autoref{eq:inf_obs} [and Eq.\ \autoref{eq:params_difference}]. Here two potential parameters are considered, namely $n=2$ and $n=4$ together with $e$-fold number $N=60$. The results are shown in Fig.\ \autoref{fig:ns_r_60} with different color-coded coupling parameter values for both methods. For $n=2$ there occurs very small difference in $n_s-r$ relations of both methods so that the curves almost coincide. That is why the differences between $n_s$ and $r$ values in both methods are also given in the right column of the figure and they both are in the order of $10^{-4}$ for $n=2$. However, for $n=4$ the difference become obvious as $\xi$ increases in observationally acceptable regime. This time the gap between $n_s$ and $r$ values are in the order of $10^{-3}$. In the GSR method for $n=4$ the curve converges as $\xi$ value increases whereas in the EF-to-JF method $\xi$ is bounded considering the observational data.

\begin{figure*}[!h]
	\centering \hspace{-3mm}
	\begin{tabular}{@{}c@{}}
		\includegraphics[width=.56\linewidth]{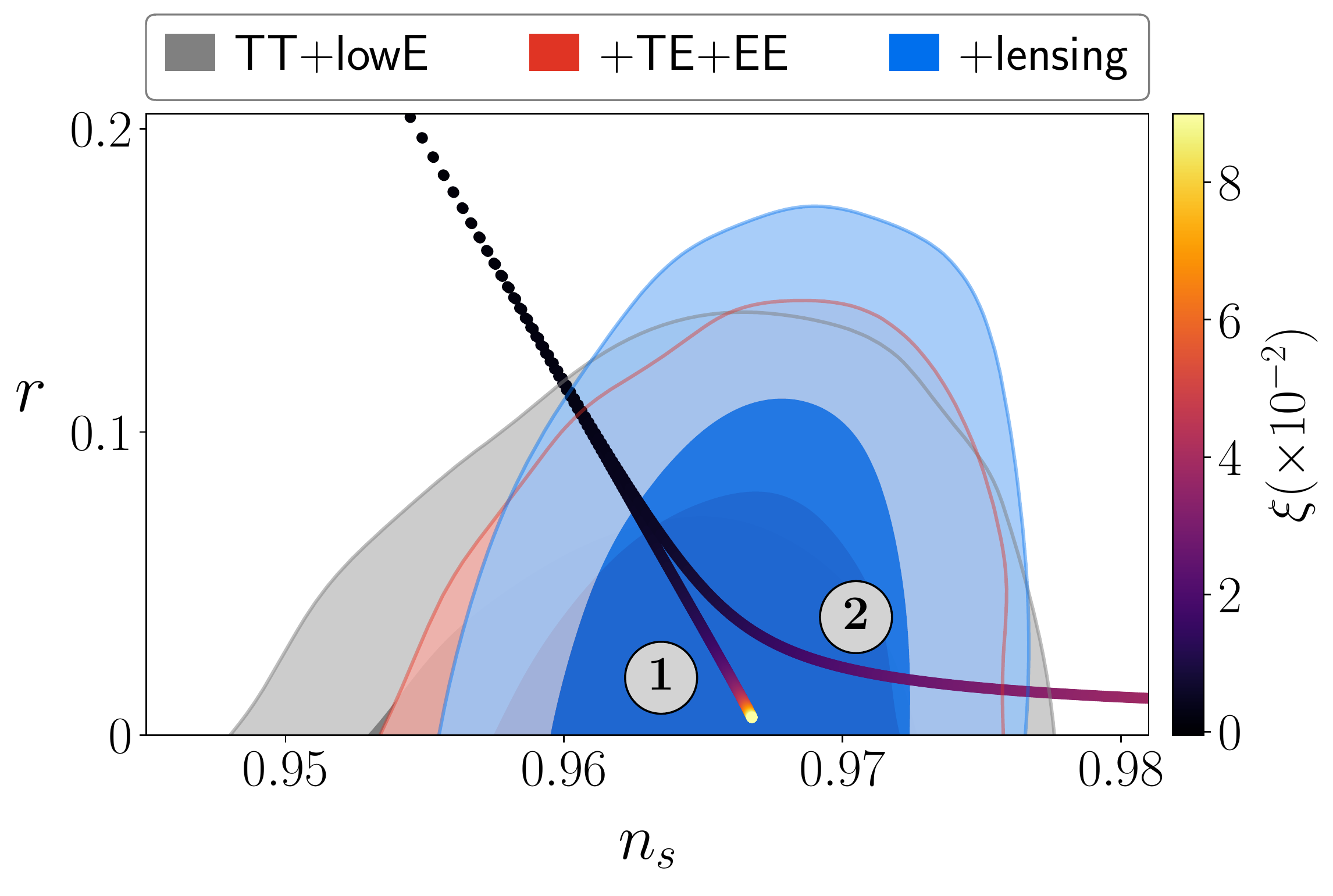} \hspace{0mm} \\
	\end{tabular}
	\begin{tabular}{@{}c@{}}
		\includegraphics[width=.42\linewidth]{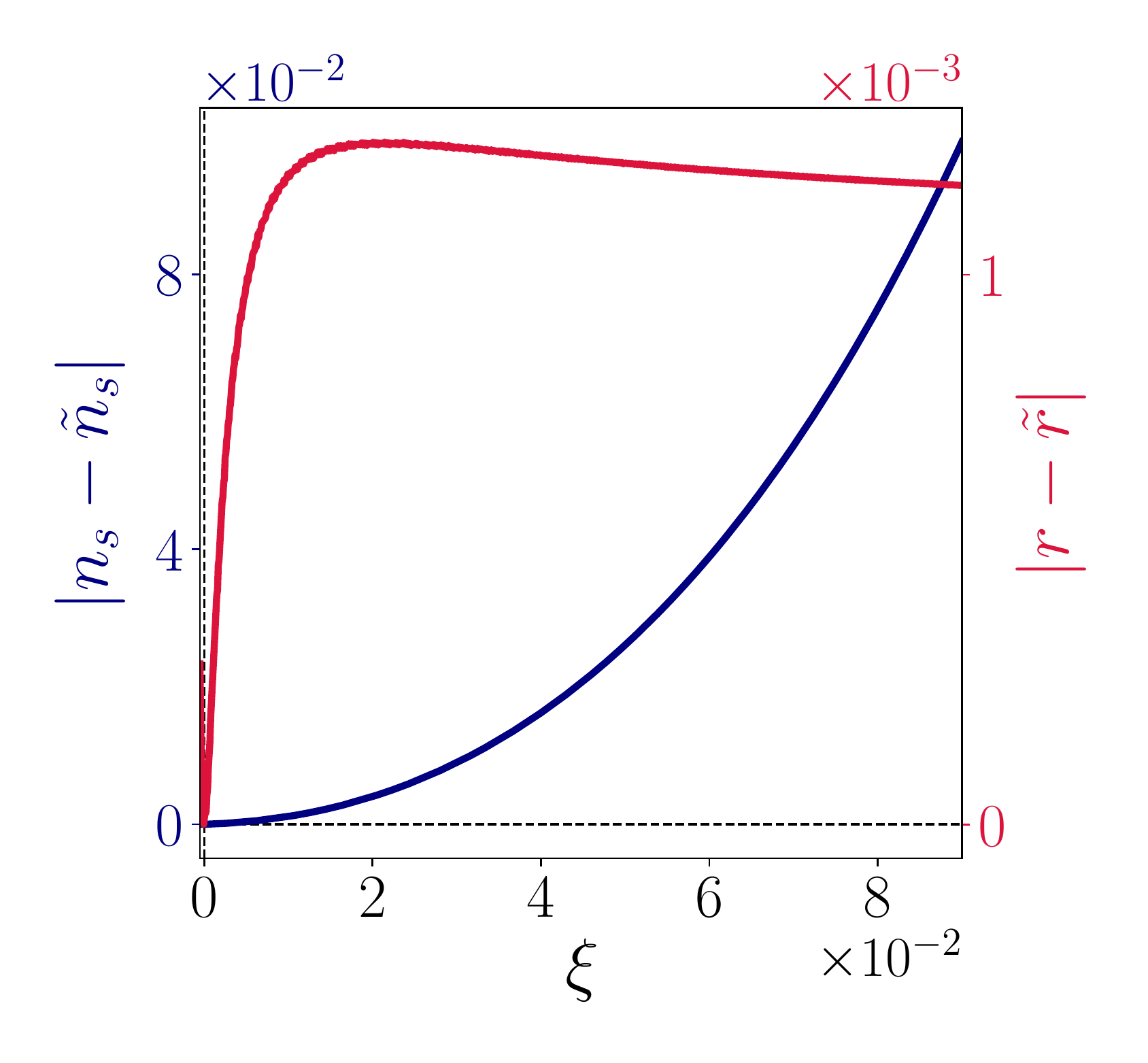} \\
	\end{tabular}
	
	\vspace{-4mm}
	
	\caption{$n_s-r$ graphs for the potential $V=V_{\!o}\,\phi^4$ with illustration of the difference of inflationary variables. Here tilde stands for the method of higher order SR. (1) and (2) in the left panel represent the solutions obtained by the GSR and the higher order SR methods, respectively.}
\label{fig:JF_GSR-SR_60}
\end{figure*}

Regarding the higher order SR approximation, for which $n_s-r$ relations is shown in Fig.\ \autoref{fig:JF_GSR-SR_60}, the analysis is not given explicitly for $n=2$ due to the fact that they do overlap and have no significant differences, in other words, the results of the GSR and the higher order SR methods coincide. However, in the case of $n=4$ the difference between outcomes of the GSR and the higher order SR methods grows as the coupling parameter $\xi$ increases within the observationally acceptable range. Therefore, it can be concluded that the higher order SR analysis matches up with the EF-to-JF approach as seen from Figs.\ \autoref{fig:ns_r_60} and \autoref{fig:JF_GSR-SR_60}. Nevertheless, this similarity can be inferred directly from the analytical expressions of inflationary variables given in the related sections of the methods.

The existence of a difference between the two approaches, namely the GSR and the EF-to-JF methods, is interesting in that in Ref.\ \cite{morris14} it has been shown that JF field equations, if expressed in terms of EF variables, agree with the EF field equations directly obtained from the EF action, provided that some consistency conditions are satisfied, and that these conditions are always met. This result implies that the two frames are, at least, mathematically equivalent which in turn implies that one can work in one frame, if there is any advantage of simplicity over the other, and then can go to the other frame. Further in Ref.\ \cite{Kaiser1995} it has been shown that the spectral indices are the same in JF and EF. The route that is followed here is in the reverse order: the approximate JF equations of motion is obtained from those of EF expressed in terms of JF variables and it is compared with the approximate equations of motion obtained directly in the JF. The difference in the results does not seem to be because of the mathematical in-equivalence of the frames but stems from the fact that the SR approximation is a very critical issue and must be applied carefully for the nonminimal coupling case. From our point of view the method that first writing the SR field equations in EF and then expressing them in terms of JF variables together with the GSR parameters seems to be safer and more precise.

The change in the scalar field equation can be expected on the ground that the conformal transformations themselves are dependent on the JF scalar field $\phi$ and that the `generalized' approximation directly in the JF cannot give exactly the same scalar equation obtained via conformal transformations from that of the EF.

\section{CONCLUSION \label{sec:conclusion}}

In this study the viability of SR approximation in STT has been studied with an example model. At first the fixed points have been investigated through the dynamical system analysis after a brief discussion on the shortcomings of the minimal coupling case. The conclusion to this formal examination is that the positions of the fixed points of the system, which depends on the values of the parameters of the model at hand, are very crucial to determine the viability of the SR approximation that is subject to the existence of the appropriate inflationary attractors.

There are two different approaches to obtain the SR equations in the JF. The first one is called the ``generalized slow-roll'' approximation which generalize the SR conditions on the scalar field in the minimal coupling case to the coupling function and those resulting conditions are applied to the system together with the original ones belonging the minimal coupling case. This whole process is implemented in the JF. The second approach, on the other hand, is expressing the equations of motion in the EF, applying SR conditions of the EF and, then, conformally transforming the resulting equations to the JF where, finally, the GSR conditions are applied. Both methods have been applied for STT in Sec.\ \autoref{sec:SR_Eqs} and it has been shown that their comparison leads to a relation given in Eq.\ \autoref{eq:params_difference} between the SR parameters and the inflationary variables. Hence, calculations of the parameters, which are to be compared with the observational data, may give different results for two methods.

As mentioned before the formal examination of STT within this study has been exemplified with a scalar field that couples to the curvature of the spacetime through the term $\xi \phi^2 \R$ together with the monomial potential in the form of $V(\phi) \propto \phi^n$ in Sec.\ \autoref{sec:example}. Phase spaces for two potential parameters, $n=2$ and $n=4$, given in Fig.\ \autoref{fig:non_minimal_phase} have been obtained with two different values of the coupling constant. As pointed out before, it has been shown that the structure of phase spaces crucially depends on parameters of the coupling function besides the exceptional case of $n=4$ in which the origin is a global stable fixed point to which all solutions converge through the inflationary attractors. Consequently, investigation of phase spaces in the beginning is necessary to see the global picture of the model and to classify the initial conditions.

In addition to the phase space analysis it has been found that Eqs.\ \autoref{eq:phi_end_sq_JF} and \autoref{eq:phi_end_sq_TF} naturally gives a constraint between the parameters of the model, $\xi$ and $n$, the result of which is illustrated in Fig.\ \autoref{fig:parameter_spaces}. In spite of the fact that the valid regions of the parameter space have some differences in both methods, it is clear that for the most relevant region, i.e. $n>0$ and $\xi>0$, as $n$ increases the coupling becomes weaker, in other words, the strong coupling is forbidden. Although this comparison is not as precise as the one coming from the observations of the inflationary variables in order to constrain the values, the outcome is important due to its convenience to compare the whole parameter space directly within the model itself. 

In Fig.\ \autoref{fig:ns_r_60} $n_s\!-\!r$ relations have been given to analyze the predictions of the model in the light of the recent observational data \cite{planck2018} again for the same potential parameters, $n=2$ and $n=4$, and the $e$-fold number $N=60$ together with the differences of $n_s$ and $r$ in terms of $\xi$ for both methods. For $n=2$ it seems that value of the coupling parameter is in the order of $10^{-3}$ whereas order of the differences between values of the inflationary variables, $n_s$ and $r$, are $10^{-4}$. Two curves obtained from two different approaches, the GSR and the EF-to-JF methods, coincide for this case. On the other hand, for the case of $n=4$, $\xi$ and the differences in $n_s$ and $r$ are in the order of $10^{-2}$ and $10^{-3}$, respectively. Two curves differ in the observable region in an obvious way. The same pattern is seen in Fig.\ \autoref{fig:JF_GSR-SR_60} as well, where the GSR method has been compared with the higher order SR approach. Since the higher order SR method is closer to the non-approximated version in comparison, this result also supports the conclusion that the EF-to-JF method is more precise as mentioned before.

\section*{Acknowledgement}

This work is dedicated to all health workers all over the world who fight with COVID-19 in the front with a great devotion.  We are also grateful to doctors, nurses, and all other workers in hospitals in Turkey for making feel us safe.

\bibliographystyle{apsrev4-1}
\bibliography{nonminimal_ref.bib}

\begin{thebibliography}{40}%
\makeatletter
\providecommand \@ifxundefined [1]{%
 \@ifx{#1\undefined}
}%
\providecommand \@ifnum [1]{%
 \ifnum #1\expandafter \@firstoftwo
 \else \expandafter \@secondoftwo
 \fi
}%
\providecommand \@ifx [1]{%
 \ifx #1\expandafter \@firstoftwo
 \else \expandafter \@secondoftwo
 \fi
}%
\providecommand \natexlab [1]{#1}%
\providecommand \enquote  [1]{``#1''}%
\providecommand \bibnamefont  [1]{#1}%
\providecommand \bibfnamefont [1]{#1}%
\providecommand \citenamefont [1]{#1}%
\providecommand \href@noop [0]{\@secondoftwo}%
\providecommand \href [0]{\begingroup \@sanitize@url \@href}%
\providecommand \@href[1]{\@@startlink{#1}\@@href}%
\providecommand \@@href[1]{\endgroup#1\@@endlink}%
\providecommand \@sanitize@url [0]{\catcode `\\12\catcode `\$12\catcode
  `\&12\catcode `\#12\catcode `\^12\catcode `\_12\catcode `\%12\relax}%
\providecommand \@@startlink[1]{}%
\providecommand \@@endlink[0]{}%
\providecommand \url  [0]{\begingroup\@sanitize@url \@url }%
\providecommand \@url [1]{\endgroup\@href {#1}{\urlprefix }}%
\providecommand \urlprefix  [0]{URL }%
\providecommand \Eprint [0]{\href }%
\providecommand \doibase [0]{http://dx.doi.org/}%
\providecommand \selectlanguage [0]{\@gobble}%
\providecommand \bibinfo  [0]{\@secondoftwo}%
\providecommand \bibfield  [0]{\@secondoftwo}%
\providecommand \translation [1]{[#1]}%
\providecommand \BibitemOpen [0]{}%
\providecommand \bibitemStop [0]{}%
\providecommand \bibitemNoStop [0]{.\EOS\space}%
\providecommand \EOS [0]{\spacefactor3000\relax}%
\providecommand \BibitemShut  [1]{\csname bibitem#1\endcsname}%
\let\auto@bib@innerbib\@empty
\bibitem [{\citenamefont {{Guth}}(1981)}]{guth81}%
  \BibitemOpen
  \bibfield  {author} {\bibinfo {author} {\bibfnamefont {A.~H.}\ \bibnamefont
  {{Guth}}},\ }\href {\doibase 10.1103/PhysRevD.23.347} {\bibfield  {journal}
  {\bibinfo  {journal} {\prd}\ }\textbf {\bibinfo {volume} {23}},\ \bibinfo
  {pages} {347} (\bibinfo {year} {1981})}\BibitemShut {NoStop}%
\bibitem [{\citenamefont {{Linde}}(1982)}]{linde82}%
  \BibitemOpen
  \bibfield  {author} {\bibinfo {author} {\bibfnamefont {A.~D.}\ \bibnamefont
  {{Linde}}},\ }\href {\doibase 10.1016/0370-2693(82)91219-9} {\bibfield
  {journal} {\bibinfo  {journal} {\plb}\ }\textbf {\bibinfo {volume} {108}},\
  \bibinfo {pages} {389} (\bibinfo {year} {1982})}\BibitemShut {NoStop}%
\bibitem [{\citenamefont {{Albrecht}}\ and\ \citenamefont
  {{Steinhardt}}(1982)}]{alb-stein82}%
  \BibitemOpen
  \bibfield  {author} {\bibinfo {author} {\bibfnamefont {A.}~\bibnamefont
  {{Albrecht}}}\ and\ \bibinfo {author} {\bibfnamefont {P.~J.}\ \bibnamefont
  {{Steinhardt}}},\ }\href {\doibase 10.1103/PhysRevLett.48.1220} {\bibfield
  {journal} {\bibinfo  {journal} {Physical Review Letters}\ }\textbf {\bibinfo
  {volume} {48}},\ \bibinfo {pages} {1220} (\bibinfo {year}
  {1982})}\BibitemShut {NoStop}%
\bibitem [{\citenamefont {Lidsey}\ \emph {et~al.}(1997)\citenamefont {Lidsey},
  \citenamefont {Liddle}, \citenamefont {Kolb}, \citenamefont {Copeland},
  \citenamefont {Barreiro},\ and\ \citenamefont {Abney}}]{RMP97}%
  \BibitemOpen
  \bibfield  {author} {\bibinfo {author} {\bibfnamefont {J.~E.}\ \bibnamefont
  {Lidsey}}, \bibinfo {author} {\bibfnamefont {A.~R.}\ \bibnamefont {Liddle}},
  \bibinfo {author} {\bibfnamefont {E.~W.}\ \bibnamefont {Kolb}}, \bibinfo
  {author} {\bibfnamefont {E.~J.}\ \bibnamefont {Copeland}}, \bibinfo {author}
  {\bibfnamefont {T.}~\bibnamefont {Barreiro}}, \ and\ \bibinfo {author}
  {\bibfnamefont {M.}~\bibnamefont {Abney}},\ }\href {\doibase
  10.1103/RevModPhys.69.373} {\bibfield  {journal} {\bibinfo  {journal} {Rev.
  Mod. Phys.}\ }\textbf {\bibinfo {volume} {69}},\ \bibinfo {pages} {373}
  (\bibinfo {year} {1997})},\ \Eprint {http://arxiv.org/abs/astro-ph/9508078}
  {\color{red}arXiv:astro-ph/9508078} \BibitemShut {NoStop}%
\bibitem [{\citenamefont {Lyth}\ and\ \citenamefont {Riotto}(1999)}]{PR99}%
  \BibitemOpen
  \bibfield  {author} {\bibinfo {author} {\bibfnamefont {D.~H.}\ \bibnamefont
  {Lyth}}\ and\ \bibinfo {author} {\bibfnamefont {A.}~\bibnamefont {Riotto}},\
  }\href {\doibase http://dx.doi.org/10.1016/S0370-1573(98)00128-8} {\bibfield
  {journal} {\bibinfo  {journal} {Physics Reports}\ }\textbf {\bibinfo {volume}
  {314}},\ \bibinfo {pages} {1 } (\bibinfo {year} {1999})}\BibitemShut
  {NoStop}%
\bibitem [{\citenamefont {Bassett}\ \emph {et~al.}(2006)\citenamefont
  {Bassett}, \citenamefont {Tsujikawa},\ and\ \citenamefont {Wands}}]{RMP06}%
  \BibitemOpen
  \bibfield  {author} {\bibinfo {author} {\bibfnamefont {B.~A.}\ \bibnamefont
  {Bassett}}, \bibinfo {author} {\bibfnamefont {S.}~\bibnamefont {Tsujikawa}},
  \ and\ \bibinfo {author} {\bibfnamefont {D.}~\bibnamefont {Wands}},\ }\href
  {\doibase 10.1103/RevModPhys.78.537} {\bibfield  {journal} {\bibinfo
  {journal} {Rev. Mod. Phys.}\ }\textbf {\bibinfo {volume} {78}},\ \bibinfo
  {pages} {537} (\bibinfo {year} {2006})}\BibitemShut {NoStop}%
\bibitem [{\citenamefont {Baumann}(2011)}]{baumann09}%
  \BibitemOpen
  \bibfield  {author} {\bibinfo {author} {\bibfnamefont {D.}~\bibnamefont
  {Baumann}},\ }in\ \href {\doibase 10.1142/9789814327183_0010} {\emph
  {\bibinfo {booktitle} {{Physics of the large and the small, TASI 09,
  proceedings of the Theoretical Advanced Study Institute in Elementary
  Particle Physics, Boulder, Colorado, USA, 1-26 June 2009}}}}\ (\bibinfo
  {year} {2011})\ pp.\ \bibinfo {pages} {523--686},\ \Eprint
  {http://arxiv.org/abs/0907.5424} {\color{red}arXiv:0907.5424} \BibitemShut
  {NoStop}%
\bibitem [{\citenamefont {{Liddle}}\ \emph {et~al.}(1994)\citenamefont
  {{Liddle}}, \citenamefont {{Parsons}},\ and\ \citenamefont
  {{Barrow}}}]{sr94}%
  \BibitemOpen
  \bibfield  {author} {\bibinfo {author} {\bibfnamefont {A.~R.}\ \bibnamefont
  {{Liddle}}}, \bibinfo {author} {\bibfnamefont {P.}~\bibnamefont {{Parsons}}},
  \ and\ \bibinfo {author} {\bibfnamefont {J.~D.}\ \bibnamefont {{Barrow}}},\
  }\href {\doibase 10.1103/PhysRevD.50.7222} {\bibfield  {journal} {\bibinfo
  {journal} {\prd}\ }\textbf {\bibinfo {volume} {50}},\ \bibinfo {pages} {7222}
  (\bibinfo {year} {1994})},\ \Eprint {http://arxiv.org/abs/astro-ph/9408015}
  {\color{red}arXiv:astro-ph/9408015} \BibitemShut {NoStop}%
\bibitem [{\citenamefont {{Faraoni}}(2000)}]{faraoni-gsr}%
  \BibitemOpen
  \bibfield  {author} {\bibinfo {author} {\bibfnamefont {V.}~\bibnamefont
  {{Faraoni}}},\ }\href {\doibase 10.1016/S0375-9601(00)00257-7} {\bibfield
  {journal} {\bibinfo  {journal} {Physics Letters A}\ }\textbf {\bibinfo
  {volume} {269}},\ \bibinfo {pages} {209} (\bibinfo {year} {2000})},\ \Eprint
  {http://arxiv.org/abs/gr-qc/0004007} {\color{red}arXiv:gr-qc/0004007}
  \BibitemShut {NoStop}%
\bibitem [{\citenamefont {{Faraoni}}(1996)}]{faraoni96}%
  \BibitemOpen
  \bibfield  {author} {\bibinfo {author} {\bibfnamefont {V.}~\bibnamefont
  {{Faraoni}}},\ }\href {\doibase 10.1103/PhysRevD.53.6813} {\bibfield
  {journal} {\bibinfo  {journal} {\prd}\ }\textbf {\bibinfo {volume} {53}},\
  \bibinfo {pages} {6813} (\bibinfo {year} {1996})},\ \Eprint
  {http://arxiv.org/abs/astro-ph/9602111} {\color{red}arXiv:astro-ph/9602111}
  \BibitemShut {NoStop}%
\bibitem [{\citenamefont {Shapiro}\ \emph {et~al.}(2015)\citenamefont
  {Shapiro}, \citenamefont {Morais~Teixeira},\ and\ \citenamefont
  {Wipf}}]{shapiro15}%
  \BibitemOpen
  \bibfield  {author} {\bibinfo {author} {\bibfnamefont {I.~L.}\ \bibnamefont
  {Shapiro}}, \bibinfo {author} {\bibfnamefont {P.}~\bibnamefont
  {Morais~Teixeira}}, \ and\ \bibinfo {author} {\bibfnamefont {A.}~\bibnamefont
  {Wipf}},\ }\href {\doibase 10.1140/epjc/s10052-015-3488-4} {\bibfield
  {journal} {\bibinfo  {journal} {Eur. Phys. J.}\ }\textbf {\bibinfo {volume}
  {C75}},\ \bibinfo {pages} {262} (\bibinfo {year} {2015})},\ \Eprint
  {http://arxiv.org/abs/1503.00874} {\color{red}arXiv:1503.00874} \BibitemShut
  {NoStop}%
\bibitem [{\citenamefont {Inagaki}\ \emph {et~al.}(2014)\citenamefont
  {Inagaki}, \citenamefont {Nakanishi},\ and\ \citenamefont
  {Odintsov}}]{odintsov14}%
  \BibitemOpen
  \bibfield  {author} {\bibinfo {author} {\bibfnamefont {T.}~\bibnamefont
  {Inagaki}}, \bibinfo {author} {\bibfnamefont {R.}~\bibnamefont {Nakanishi}},
  \ and\ \bibinfo {author} {\bibfnamefont {S.~D.}\ \bibnamefont {Odintsov}},\
  }\href {\doibase 10.1007/s10509-014-2108-3} {\bibfield  {journal} {\bibinfo
  {journal} {Astrophys. Space Sci.}\ }\textbf {\bibinfo {volume} {354}},\
  \bibinfo {pages} {2108} (\bibinfo {year} {2014})},\ \Eprint
  {http://arxiv.org/abs/1408.1270} {\color{red}arXiv:1408.1270} \BibitemShut
  {NoStop}%
\bibitem [{\citenamefont {Inagaki}\ \emph {et~al.}(2015)\citenamefont
  {Inagaki}, \citenamefont {Nakanishi},\ and\ \citenamefont
  {Odintsov}}]{odintsov15}%
  \BibitemOpen
  \bibfield  {author} {\bibinfo {author} {\bibfnamefont {T.}~\bibnamefont
  {Inagaki}}, \bibinfo {author} {\bibfnamefont {R.}~\bibnamefont {Nakanishi}},
  \ and\ \bibinfo {author} {\bibfnamefont {S.~D.}\ \bibnamefont {Odintsov}},\
  }\href {\doibase 10.1016/j.physletb.2015.04.038} {\bibfield  {journal}
  {\bibinfo  {journal} {Phys. Lett.}\ }\textbf {\bibinfo {volume} {B745}},\
  \bibinfo {pages} {105} (\bibinfo {year} {2015})},\ \Eprint
  {http://arxiv.org/abs/1502.06301} {\color{red}arXiv:1502.06301} \BibitemShut
  {NoStop}%
\bibitem [{\citenamefont {Myrzakul}\ \emph {et~al.}(2016)\citenamefont
  {Myrzakul}, \citenamefont {Myrzakulov},\ and\ \citenamefont
  {Sebastiani}}]{myrzakul15}%
  \BibitemOpen
  \bibfield  {author} {\bibinfo {author} {\bibfnamefont {S.}~\bibnamefont
  {Myrzakul}}, \bibinfo {author} {\bibfnamefont {R.}~\bibnamefont
  {Myrzakulov}}, \ and\ \bibinfo {author} {\bibfnamefont {L.}~\bibnamefont
  {Sebastiani}},\ }\href {\doibase 10.1142/S0218271816500413} {\bibfield
  {journal} {\bibinfo  {journal} {Int. J. Mod. Phys.}\ }\textbf {\bibinfo
  {volume} {D25}},\ \bibinfo {pages} {1650041} (\bibinfo {year} {2016})},\
  \Eprint {http://arxiv.org/abs/1509.07021} {\color{red}arXiv:1509.07021}
  \BibitemShut {NoStop}%
\bibitem [{\citenamefont {Myrzakulov}\ \emph {et~al.}(2015)\citenamefont
  {Myrzakulov}, \citenamefont {Sebastiani},\ and\ \citenamefont
  {Vagnozzi}}]{myrzakul15a}%
  \BibitemOpen
  \bibfield  {author} {\bibinfo {author} {\bibfnamefont {R.}~\bibnamefont
  {Myrzakulov}}, \bibinfo {author} {\bibfnamefont {L.}~\bibnamefont
  {Sebastiani}}, \ and\ \bibinfo {author} {\bibfnamefont {S.}~\bibnamefont
  {Vagnozzi}},\ }\href {\doibase 10.1140/epjc/s10052-015-3672-6} {\bibfield
  {journal} {\bibinfo  {journal} {Eur. Phys. J.}\ }\textbf {\bibinfo {volume}
  {C75}},\ \bibinfo {pages} {444} (\bibinfo {year} {2015})},\ \Eprint
  {http://arxiv.org/abs/1504.07984} {\color{red}arXiv:1504.07984} \BibitemShut
  {NoStop}%
\bibitem [{\citenamefont {Zubair}\ and\ \citenamefont
  {Kousar}(2017)}]{zubair17}%
  \BibitemOpen
  \bibfield  {author} {\bibinfo {author} {\bibfnamefont {M.}~\bibnamefont
  {Zubair}}\ and\ \bibinfo {author} {\bibfnamefont {F.}~\bibnamefont
  {Kousar}},\ }\href {\doibase 10.1139/cjp-2017-0075} {\bibfield  {journal}
  {\bibinfo  {journal} {Can. J. Phys.}\ }\textbf {\bibinfo {volume} {95}},\
  \bibinfo {pages} {1074} (\bibinfo {year} {2017})}\BibitemShut {NoStop}%
\bibitem [{\citenamefont {van~de Bruck}\ and\ \citenamefont
  {Longden}(2016)}]{carsten16}%
  \BibitemOpen
  \bibfield  {author} {\bibinfo {author} {\bibfnamefont {C.}~\bibnamefont
  {van~de Bruck}}\ and\ \bibinfo {author} {\bibfnamefont {C.}~\bibnamefont
  {Longden}},\ }\href {\doibase 10.1103/PhysRevD.93.063519} {\bibfield
  {journal} {\bibinfo  {journal} {Phys. Rev.}\ }\textbf {\bibinfo {volume}
  {D93}},\ \bibinfo {pages} {063519} (\bibinfo {year} {2016})},\ \Eprint
  {http://arxiv.org/abs/1512.04768} {\color{red}arXiv:1512.04768} \BibitemShut
  {NoStop}%
\bibitem [{\citenamefont {He}\ and\ \citenamefont {Piao}(2019)}]{he19}%
  \BibitemOpen
  \bibfield  {author} {\bibinfo {author} {\bibfnamefont {Y.-L.}\ \bibnamefont
  {He}}\ and\ \bibinfo {author} {\bibfnamefont {Y.-S.}\ \bibnamefont {Piao}},\
  }\href {\doibase 10.1103/PhysRevD.99.083511} {\bibfield  {journal} {\bibinfo
  {journal} {Phys. Rev.}\ }\textbf {\bibinfo {volume} {D99}},\ \bibinfo {pages}
  {083511} (\bibinfo {year} {2019})},\ \Eprint
  {http://arxiv.org/abs/1810.00202} {\color{red}arXiv:1810.00202} \BibitemShut
  {NoStop}%
\bibitem [{\citenamefont {Granda}\ and\ \citenamefont
  {Jimenez}(2019{\natexlab{a}})}]{granda1}%
  \BibitemOpen
  \bibfield  {author} {\bibinfo {author} {\bibfnamefont {L.~N.}\ \bibnamefont
  {Granda}}\ and\ \bibinfo {author} {\bibfnamefont {D.~F.}\ \bibnamefont
  {Jimenez}},\ }\href@noop {} {\  (\bibinfo {year} {2019}{\natexlab{a}})},\
  \Eprint {http://arxiv.org/abs/1910.11289} {\color{red}arXiv:1910.11289}
  \BibitemShut {NoStop}%
\bibitem [{\citenamefont {Granda}\ and\ \citenamefont
  {Jimenez}(2019{\natexlab{b}})}]{granda2}%
  \BibitemOpen
  \bibfield  {author} {\bibinfo {author} {\bibfnamefont {L.~N.}\ \bibnamefont
  {Granda}}\ and\ \bibinfo {author} {\bibfnamefont {D.~F.}\ \bibnamefont
  {Jimenez}},\ }\href {\doibase 10.1140/epjc/s10052-019-7289-z} {\bibfield
  {journal} {\bibinfo  {journal} {Eur. Phys. J.}\ }\textbf {\bibinfo {volume}
  {C79}},\ \bibinfo {pages} {772} (\bibinfo {year} {2019}{\natexlab{b}})},\
  \Eprint {http://arxiv.org/abs/1907.06806} {\color{red}arXiv:1907.06806}
  \BibitemShut {NoStop}%
\bibitem [{\citenamefont {Granda}\ \emph {et~al.}(2019)\citenamefont {Granda},
  \citenamefont {Jimenez},\ and\ \citenamefont {Cardona}}]{granda3}%
  \BibitemOpen
  \bibfield  {author} {\bibinfo {author} {\bibfnamefont {L.~N.}\ \bibnamefont
  {Granda}}, \bibinfo {author} {\bibfnamefont {D.~F.}\ \bibnamefont {Jimenez}},
  \ and\ \bibinfo {author} {\bibfnamefont {W.}~\bibnamefont {Cardona}},\
  }\href@noop {} {\  (\bibinfo {year} {2019})},\ \Eprint
  {http://arxiv.org/abs/1911.02901} {\color{red}arXiv:1911.02901} \BibitemShut
  {NoStop}%
\bibitem [{\citenamefont {{Takahashi}}\ \emph {et~al.}(2020)\citenamefont
  {{Takahashi}}, \citenamefont {{Tenkanen}},\ and\ \citenamefont
  {{Yokoyama}}}]{taka2003}%
  \BibitemOpen
  \bibfield  {author} {\bibinfo {author} {\bibfnamefont {T.}~\bibnamefont
  {{Takahashi}}}, \bibinfo {author} {\bibfnamefont {T.}~\bibnamefont
  {{Tenkanen}}}, \ and\ \bibinfo {author} {\bibfnamefont {S.}~\bibnamefont
  {{Yokoyama}}},\ }\href@noop {} {\  (\bibinfo {year} {2020})},\ \Eprint
  {http://arxiv.org/abs/2003.10203} {\color{red}arXiv:2003.10203} \BibitemShut
  {NoStop}%
\bibitem [{\citenamefont {Akrami}\ \emph {et~al.}(2018)\citenamefont {Akrami}
  \emph {et~al.}}]{planck2018}%
  \BibitemOpen
  \bibfield  {author} {\bibinfo {author} {\bibfnamefont {Y.}~\bibnamefont
  {Akrami}} \emph {et~al.} (\bibinfo {collaboration} {Planck Collaboration}),\
  }\href@noop {} {\  (\bibinfo {year} {2018})},\ \Eprint
  {http://arxiv.org/abs/1807.06211} {\color{red}arXiv:1807.06211} \BibitemShut
  {NoStop}%
\bibitem [{\citenamefont {Lewis}(2019)}]{Lewis2019_getdist}%
  \BibitemOpen
  \bibfield  {author} {\bibinfo {author} {\bibfnamefont {A.}~\bibnamefont
  {Lewis}},\ }\href {https://getdist.readthedocs.io} {\  (\bibinfo {year}
  {2019})},\ \Eprint {http://arxiv.org/abs/1910.13970}
  {\color{red}arXiv:1910.13970} \BibitemShut {NoStop}%
\bibitem [{\citenamefont {Chiba}\ \emph {et~al.}(2020)\citenamefont {Chiba},
  \citenamefont {Chibana},\ and\ \citenamefont {Yamaguchi}}]{chiba2003}%
  \BibitemOpen
  \bibfield  {author} {\bibinfo {author} {\bibfnamefont {T.}~\bibnamefont
  {Chiba}}, \bibinfo {author} {\bibfnamefont {F.}~\bibnamefont {Chibana}}, \
  and\ \bibinfo {author} {\bibfnamefont {M.}~\bibnamefont {Yamaguchi}},\
  }\href@noop {} {\  (\bibinfo {year} {2020})},\ \Eprint
  {http://arxiv.org/abs/2003.10633} {\color{red}arXiv:2003.10633} \BibitemShut
  {NoStop}%
\bibitem [{\citenamefont {Chiba}\ and\ \citenamefont
  {Yamaguchi}(2013)}]{chiba13}%
  \BibitemOpen
  \bibfield  {author} {\bibinfo {author} {\bibfnamefont {T.}~\bibnamefont
  {Chiba}}\ and\ \bibinfo {author} {\bibfnamefont {M.}~\bibnamefont
  {Yamaguchi}},\ }\href {\doibase 10.1088/1475-7516/2013/10/040} {\bibfield
  {journal} {\bibinfo  {journal} {JCAP}\ }\textbf {\bibinfo {volume} {1310}},\
  \bibinfo {pages} {040} (\bibinfo {year} {2013})},\ \Eprint
  {http://arxiv.org/abs/1308.1142} {\color{red}arXiv:1308.1142} \BibitemShut
  {NoStop}%
\bibitem [{\citenamefont {Deruelle}\ and\ \citenamefont
  {Sasaki}(2011)}]{sasaki10}%
  \BibitemOpen
  \bibfield  {author} {\bibinfo {author} {\bibfnamefont {N.}~\bibnamefont
  {Deruelle}}\ and\ \bibinfo {author} {\bibfnamefont {M.}~\bibnamefont
  {Sasaki}},\ }\bibfield  {booktitle} {\emph {\bibinfo {booktitle}
  {{Proceedings, Cosmology, the Quantum Vacuum, and Zeta Functions: Bellaterra,
  Barcelona, Spain, March 8-10, 2010}}},\ }\href {\doibase
  10.1007/978-3-642-19760-4_23} {\bibfield  {journal} {\bibinfo  {journal}
  {Springer Proc. Phys.}\ }\textbf {\bibinfo {volume} {137}},\ \bibinfo {pages}
  {247} (\bibinfo {year} {2011})},\ \Eprint {http://arxiv.org/abs/1007.3563}
  {\color{red}arXiv:1007.3563} \BibitemShut {NoStop}%
\bibitem [{\citenamefont {Gong}\ \emph {et~al.}(2011)\citenamefont {Gong},
  \citenamefont {Hwang}, \citenamefont {Park}, \citenamefont {Sasaki},\ and\
  \citenamefont {Song}}]{sasaki11}%
  \BibitemOpen
  \bibfield  {author} {\bibinfo {author} {\bibfnamefont {J.-O.}\ \bibnamefont
  {Gong}}, \bibinfo {author} {\bibfnamefont {J.-c.}\ \bibnamefont {Hwang}},
  \bibinfo {author} {\bibfnamefont {W.-I.}\ \bibnamefont {Park}}, \bibinfo
  {author} {\bibfnamefont {M.}~\bibnamefont {Sasaki}}, \ and\ \bibinfo {author}
  {\bibfnamefont {Y.-S.}\ \bibnamefont {Song}},\ }\href {\doibase
  10.1088/1475-7516/2011/09/023} {\bibfield  {journal} {\bibinfo  {journal}
  {JCAP}\ }\textbf {\bibinfo {volume} {1109}},\ \bibinfo {pages} {023}
  (\bibinfo {year} {2011})},\ \Eprint {http://arxiv.org/abs/1107.1840}
  {\color{red}arXiv:1107.1840} \BibitemShut {NoStop}%
\bibitem [{\citenamefont {Kamenshchik}\ and\ \citenamefont
  {Steinwachs}(2015)}]{kamen15}%
  \BibitemOpen
  \bibfield  {author} {\bibinfo {author} {\bibfnamefont {A.~{\relax Yu}.}\
  \bibnamefont {Kamenshchik}}\ and\ \bibinfo {author} {\bibfnamefont {C.~F.}\
  \bibnamefont {Steinwachs}},\ }\href {\doibase 10.1103/PhysRevD.91.084033}
  {\bibfield  {journal} {\bibinfo  {journal} {Phys. Rev.}\ }\textbf {\bibinfo
  {volume} {D91}},\ \bibinfo {pages} {084033} (\bibinfo {year} {2015})},\
  \Eprint {http://arxiv.org/abs/1408.5769} {\color{red}arXiv:1408.5769}
  \BibitemShut {NoStop}%
\bibitem [{\citenamefont {Ohta}(2018)}]{ohta17}%
  \BibitemOpen
  \bibfield  {author} {\bibinfo {author} {\bibfnamefont {N.}~\bibnamefont
  {Ohta}},\ }\href {\doibase 10.1093/ptep/pty008} {\bibfield  {journal}
  {\bibinfo  {journal} {PTEP}\ }\textbf {\bibinfo {volume} {2018}},\ \bibinfo
  {pages} {033B02} (\bibinfo {year} {2018})},\ \Eprint
  {http://arxiv.org/abs/1712.05175} {\color{red}arXiv:1712.05175} \BibitemShut
  {NoStop}%
\bibitem [{\citenamefont {Ruf}\ and\ \citenamefont {Steinwachs}(2018)}]{ruf18}%
  \BibitemOpen
  \bibfield  {author} {\bibinfo {author} {\bibfnamefont {M.~S.}\ \bibnamefont
  {Ruf}}\ and\ \bibinfo {author} {\bibfnamefont {C.~F.}\ \bibnamefont
  {Steinwachs}},\ }\href {\doibase 10.1103/PhysRevD.97.044050} {\bibfield
  {journal} {\bibinfo  {journal} {Phys. Rev.}\ }\textbf {\bibinfo {volume}
  {D97}},\ \bibinfo {pages} {044050} (\bibinfo {year} {2018})},\ \Eprint
  {http://arxiv.org/abs/1711.07486} {\color{red}arXiv:1711.07486} \BibitemShut
  {NoStop}%
\bibitem [{\citenamefont {Kuusk}\ \emph {et~al.}(2016)\citenamefont {Kuusk},
  \citenamefont {Rünkla}, \citenamefont {Saal},\ and\ \citenamefont
  {Vilson}}]{kuusk16}%
  \BibitemOpen
  \bibfield  {author} {\bibinfo {author} {\bibfnamefont {P.}~\bibnamefont
  {Kuusk}}, \bibinfo {author} {\bibfnamefont {M.}~\bibnamefont {Rünkla}},
  \bibinfo {author} {\bibfnamefont {M.}~\bibnamefont {Saal}}, \ and\ \bibinfo
  {author} {\bibfnamefont {O.}~\bibnamefont {Vilson}},\ }\href {\doibase
  10.1088/0264-9381/33/19/195008} {\bibfield  {journal} {\bibinfo  {journal}
  {Class. Quant. Grav.}\ }\textbf {\bibinfo {volume} {33}},\ \bibinfo {pages}
  {195008} (\bibinfo {year} {2016})},\ \Eprint
  {http://arxiv.org/abs/1605.07033} {\color{red}arXiv:1605.07033} \BibitemShut
  {NoStop}%
\bibitem [{\citenamefont {Järv}\ \emph {et~al.}(2017)\citenamefont {Järv},
  \citenamefont {Kannike}, \citenamefont {Marzola}, \citenamefont {Racioppi},
  \citenamefont {Raidal}, \citenamefont {Rünkla}, \citenamefont {Saal},\ and\
  \citenamefont {Veermäe}}]{kannike17}%
  \BibitemOpen
  \bibfield  {author} {\bibinfo {author} {\bibfnamefont {L.}~\bibnamefont
  {Järv}}, \bibinfo {author} {\bibfnamefont {K.}~\bibnamefont {Kannike}},
  \bibinfo {author} {\bibfnamefont {L.}~\bibnamefont {Marzola}}, \bibinfo
  {author} {\bibfnamefont {A.}~\bibnamefont {Racioppi}}, \bibinfo {author}
  {\bibfnamefont {M.}~\bibnamefont {Raidal}}, \bibinfo {author} {\bibfnamefont
  {M.}~\bibnamefont {Rünkla}}, \bibinfo {author} {\bibfnamefont
  {M.}~\bibnamefont {Saal}}, \ and\ \bibinfo {author} {\bibfnamefont
  {H.}~\bibnamefont {Veermäe}},\ }\href {\doibase
  10.1103/PhysRevLett.118.151302} {\bibfield  {journal} {\bibinfo  {journal}
  {Phys. Rev. Lett.}\ }\textbf {\bibinfo {volume} {118}},\ \bibinfo {pages}
  {151302} (\bibinfo {year} {2017})},\ \Eprint
  {http://arxiv.org/abs/1612.06863} {\color{red}arXiv:1612.06863} \BibitemShut
  {NoStop}%
\bibitem [{\citenamefont {Karam}\ \emph {et~al.}(2017)\citenamefont {Karam},
  \citenamefont {Pappas},\ and\ \citenamefont {Tamvakis}}]{tamvakis17}%
  \BibitemOpen
  \bibfield  {author} {\bibinfo {author} {\bibfnamefont {A.}~\bibnamefont
  {Karam}}, \bibinfo {author} {\bibfnamefont {T.}~\bibnamefont {Pappas}}, \
  and\ \bibinfo {author} {\bibfnamefont {K.}~\bibnamefont {Tamvakis}},\ }\href
  {\doibase 10.1103/PhysRevD.96.064036} {\bibfield  {journal} {\bibinfo
  {journal} {Phys. Rev.}\ }\textbf {\bibinfo {volume} {D96}},\ \bibinfo {pages}
  {064036} (\bibinfo {year} {2017})},\ \Eprint
  {http://arxiv.org/abs/1707.00984} {\color{red}arXiv:1707.00984} \BibitemShut
  {NoStop}%
\bibitem [{\citenamefont {{Morris}}(2001)}]{morris01}%
  \BibitemOpen
  \bibfield  {author} {\bibinfo {author} {\bibfnamefont {J.~R.}\ \bibnamefont
  {{Morris}}},\ }\href {\doibase 10.1088/0264-9381/18/15/311} {\bibfield
  {journal} {\bibinfo  {journal} {Classical and Quantum Gravity}\ }\textbf
  {\bibinfo {volume} {18}},\ \bibinfo {pages} {2977} (\bibinfo {year}
  {2001})},\ \Eprint {http://arxiv.org/abs/gr-qc/0106022}
  {\color{red}arXiv:gr-qc/0106022} \BibitemShut {NoStop}%
\bibitem [{\citenamefont {{Torres}}(1997)}]{torres97}%
  \BibitemOpen
  \bibfield  {author} {\bibinfo {author} {\bibfnamefont {D.~F.}\ \bibnamefont
  {{Torres}}},\ }\href {\doibase 10.1016/S0375-9601(96)00835-3} {\bibfield
  {journal} {\bibinfo  {journal} {\pla}\ }\textbf {\bibinfo {volume} {225}},\
  \bibinfo {pages} {13} (\bibinfo {year} {1997})},\ \Eprint
  {http://arxiv.org/abs/gr-qc/9610021} {\color{red}arXiv:gr-qc/9610021}
  \BibitemShut {NoStop}%
\bibitem [{\citenamefont {Granda}\ and\ \citenamefont
  {Jimenez}(2019{\natexlab{c}})}]{granda2019}%
  \BibitemOpen
  \bibfield  {author} {\bibinfo {author} {\bibfnamefont {L.~N.}\ \bibnamefont
  {Granda}}\ and\ \bibinfo {author} {\bibfnamefont {D.~F.}\ \bibnamefont
  {Jimenez}},\ }\href {\doibase 10.1088/1475-7516/2019/09/007} {\bibfield
  {journal} {\bibinfo  {journal} {JCAP}\ }\textbf {\bibinfo {volume} {1909}},\
  \bibinfo {pages} {007} (\bibinfo {year} {2019}{\natexlab{c}})},\ \Eprint
  {http://arxiv.org/abs/1905.08349} {\color{red}arXiv:1905.08349} \BibitemShut
  {NoStop}%
\bibitem [{\citenamefont {Amendola}\ \emph {et~al.}(1990)\citenamefont
  {Amendola}, \citenamefont {Litterio},\ and\ \citenamefont
  {Occhionero}}]{Amendola1990}%
  \BibitemOpen
  \bibfield  {author} {\bibinfo {author} {\bibfnamefont {L.}~\bibnamefont
  {Amendola}}, \bibinfo {author} {\bibfnamefont {M.}~\bibnamefont {Litterio}},
  \ and\ \bibinfo {author} {\bibfnamefont {F.}~\bibnamefont {Occhionero}},\
  }\href {\doibase 10.1142/S0217751X90001653} {\bibfield  {journal} {\bibinfo
  {journal} {Int. J. Mod. Phys.}\ }\textbf {\bibinfo {volume} {A5}},\ \bibinfo
  {pages} {3861} (\bibinfo {year} {1990})}\BibitemShut {NoStop}%
\bibitem [{\citenamefont {{Morris}}(2014)}]{morris14}%
  \BibitemOpen
  \bibfield  {author} {\bibinfo {author} {\bibfnamefont {J.~R.}\ \bibnamefont
  {{Morris}}},\ }\href {\doibase 10.1103/PhysRevD.90.107501} {\bibfield
  {journal} {\bibinfo  {journal} {\prd}\ }\textbf {\bibinfo {volume} {90}},\
  \bibinfo {eid} {107501} (\bibinfo {year} {2014})},\ \Eprint
  {http://arxiv.org/abs/1411.1311} {\color{red}arXiv:1411.1311} \BibitemShut
  {NoStop}%
\bibitem [{\citenamefont {Kaiser}(1995)}]{Kaiser1995}%
  \BibitemOpen
  \bibfield  {author} {\bibinfo {author} {\bibfnamefont {D.~I.}\ \bibnamefont
  {Kaiser}},\ }\href@noop {} {\bibfield  {journal} {\bibinfo  {journal}
  {Submitted to: Phys. Lett. B}\ } (\bibinfo {year} {1995})},\ \Eprint
  {http://arxiv.org/abs/astro-ph/9507048} {\color{red}arXiv:astro-ph/9507048}
  \BibitemShut {NoStop}%
\end{thebibliography}%

\end{document}